\documentclass[12pt]{article}

\usepackage[a4paper,margin=1in]{geometry}
\usepackage{amsmath,amssymb,amsfonts,bm}
\usepackage{graphicx}
\usepackage{booktabs}
\usepackage{multirow}
\usepackage{siunitx}
\usepackage{subcaption}
\usepackage{hyphenat}
\usepackage{algorithm}
\usepackage{algpseudocode}
\usepackage{enumitem}
\usepackage{microtype}
\usepackage{caption}
\usepackage{xcolor}
\usepackage{float}
\usepackage[hidelinks]{hyperref}
\usepackage{tikz}
\usetikzlibrary{arrows.meta,positioning,fit,calc}
\usepackage{autonum}
\definecolor{Reviewer1}{rgb}{0,0,0.9}
\definecolor{Reviewer2}{rgb}{0.8,0,0}
\definecolor{Reviewer4}{rgb}{0,0.5,0}

\makeatletter
\providecommand{\caption@x@label}[2]{#1{#2}} 
\makeatother

\newcommand{\R}{\mathbb{R}}
\newcommand{\dd}{\mathrm{d}}
\newcommand{\Tt}{\mathcal{T}} 

\title{Nonlinear Dynamic Modeling and Receding-Horizon NMPC of an Electric Unicycle on a Tensioned Cable}
\author{%
  Yousef Sweiti~~Jasem Tamimi\textsuperscript{*}   \\
  \small Department of Mechanical Engineering,\\ \small Palestine Polytechnic University, Hebron, Palestine\\
  \small \textsuperscript{*}Corresponding author: jtamimi@ppu.edu%
}
\date{}

\begin{document}
\maketitle

\begin{abstract}
Traversing flexible slender structures involves strong bidirectional coupling between vehicle motion and structural vibration caused by a moving contact point.
This paper presents a nonlinear modeling-and-control framework for an electric unicycle traversing a tensioned flexible cable.
The cable is modeled using a tension-dominated finite-element formulation retaining vertical and lateral transverse dynamics, and is coupled to the unicycle roll and pitch dynamics through moving-contact kinematics.
A modal truncation yields a control-oriented reduced-order model that preserves the nonlinear vehicle--cable coupling.
Based on this model, a constrained nonlinear model predictive controller is formulated to regulate traversal, balance, actuator limits, and cable-span constraints.
To reduce the computational burden associated with moving-contact interpolation, a frozen spatial contact-point interpolation strategy is introduced for the NMPC prediction model.
Numerical studies under nominal, disturbed, constrained, and model-mismatch scenarios show accurate traversal, roll and pitch stabilization, attenuation of cable vibrations, and constraint-consistent closed-loop behavior.
{The frozen-interpolation formulation reduces and regularizes NMPC computation times relative to the non-frozen formulation, supporting its use as a computationally tractable approach for finite-element-based moving-contact vehicle--structure interaction systems.}
\end{abstract}

\section{Introduction}\label{sec:introduction}

{Robotic traversal of flexible slender structures is challenging because the vehicle
motion and structural vibration are coupled through a continuously moving contact point.
In a unicycle--on--cable system, the wheel contact location migrates along the cable span,
so the local cable deflection and slope directly affect the unicycle kinematics, inertial
loading, and roll--pitch balance dynamics.
At the same time, wheel actuation and vehicle motion inject moving loads into the cable and
excite distributed transverse vibrations.
The resulting vehicle--structure interaction is particularly demanding for narrow-support
vehicles, where tracking errors, cable oscillations, actuator limits, and external
disturbances can compromise both traversal accuracy and attitude stability.
Therefore, safe cable traversal requires a control-oriented model and controller that
explicitly account for nonlinear vehicle--cable coupling, moving-contact constraints,
actuator limitations, and balance requirements.}

Vehicle--structure interaction problems have been extensively investigated for moving loads
on beams, cables, and bridges, where the structural response depends on the instantaneous
vehicle position and the associated contact or moving-load forces.
For taut cables, classical formulations commonly neglect bending stiffness and describe the
transverse response using tension-dominated wave equations
\cite{Zaman1996,Au2001,Yin2011}.
Finite-element discretization of these models provides an accurate finite-dimensional
description, but the resulting dynamics may become computationally demanding for online
control when the contact point moves continuously across finite elements.
Modal reduction alleviates this burden by projecting the cable dynamics onto a truncated set
of dominant modes
\cite{Karoumi2000,Sun2011,Hurel2019}.

{Despite these advances, most cable--vehicle interaction studies focus on vertical
structural response or open-loop dynamic analysis, and therefore do not directly address the
roll-stability requirements of narrow-support vehicles.}
In particular, structural models restricted to vertical deflection
\cite{Yin2011,Au2001,Karoumi2000} cannot capture the lateral cable motion that directly
affects roll dynamics in a unicycle--on--cable system.
Hurel et al.~\cite{Hurel2019} consider both lateral and vertical dynamics in a bi-cable
ropeway system; however, the repeated update of modal bases at each time step increases the
online computational burden and is not directly suited to constrained nonlinear
receding-horizon control.
{These limitations motivate a control-oriented cable model that retains both vertical
and lateral transverse dynamics while remaining compatible with online NMPC.}

Model predictive control (MPC) is well suited to flexible and distributed-parameter
systems because it can handle constraints, nonlinear dynamics, and multiple performance
objectives within a receding-horizon framework
\cite{Rawlings2017}.
Artola et al.~\cite{Artola2021} developed a modal-based nonlinear MPC formulation for
geometrically exact three-dimensional flexible beams, including the treatment of modal
truncation effects and uncontrollable modes.
Related control strategies for flexible systems, such as energy-shaping and adaptive
disturbance compensation
\cite{Franco2018}, improve robustness in the presence of structural flexibility, but do not
formulate a constrained receding-horizon NMPC problem for a moving-contact cable--vehicle
system.
In parallel, NMPC implementations based on CasADi~\cite{Andersson2019} and IPOPT
\cite{Wachter2006} have been demonstrated for complex vehicle systems
\cite{Moradi2022,vanLookerenCampagne2019}.

{However, existing NMPC and flexible-system control studies generally do not address the
specific computational structure introduced by vehicle--structure interaction with a moving
contact point. In particular, the position-dependent interpolation and slope operators that
couple the vehicle to a finite-element cable model are absent from conventional flexible-beam
NMPC and vehicle-control formulations.}
{Consequently, the integration of modal cable dynamics, nonlinear unicycle balance,
moving-contact interpolation, and constrained NMPC remains insufficiently developed for
online closed-loop control.}

A central computational challenge in VSI-based NMPC arises from the position-dependent
nature of the moving-contact operators.
As the vehicle advances along the span, the interpolation and slope operators that map
modal cable coordinates to contact-point deflections and slopes vary with the axial position
and must be updated consistently within the prediction model.
In finite-element-based formulations, this dependence may introduce repeated element
identification, operator evaluation, and contact-coupling updates inside the nonlinear
optimization loop.

Tsao et al.~\cite{Tsao2001} proposed a control-oriented partitioning that separates
time-invariant structural and vehicle dynamics from time-varying moving-contact components,
leading to an LPV representation suitable for gain-scheduling control.
Karoumi~\cite{Karoumi1998} used linear interpolation functions with coefficients updated
at each time step for vehicle--bridge interaction, but the interpolation operators were not
frozen across a nonlinear prediction horizon.
Recent linear time-varying formulations based on modal decomposition, such as
Zhao et al.~\cite{Zhao2023}, similarly represent the moving contact through
time-dependent coupling terms acting on an otherwise time-invariant structural model.

{Although LPV, linear time-varying, and frozen-time receding-horizon formulations provide
useful control-oriented representations, they do not directly address the combination of
nonlinear unicycle balance dynamics, modal cable coordinates, state-dependent finite-element
contact interpolation, and constrained nonlinear receding-horizon optimization considered
here.}
Frozen-time RHC freezes temporally varying system matrices to recover time-invariant Riccati
or optimization structures
\cite{Oh2010,Oh2011}, whereas the present problem involves spatially varying interpolation
and slope operators governed by the vehicle position along the cable span.
Related frozen-parameter concepts have also been used in LPV--MPC formulations based on
scheduling variables~\cite{Floch2025} and model-predictive schemes involving fixed or
pretrained aerodynamic models~\cite{shu2025model}.
{To the authors' knowledge, freezing spatial contact-point interpolation operators over
the NMPC prediction horizon has not been explicitly treated for nonlinear FEM-based
moving-contact cable--vehicle systems.}
This motivates the full nonlinear NMPC formulation with frozen spatial contact-point
interpolation developed in this work, rather than an LPV--MPC formulation based on repeated
linearization or dense scheduling grids
\cite{Tsao2001,Floch2025}.

Beyond computational issues, most cable--vehicle interaction studies focus on open-loop
dynamic response analysis
\cite{Au2001,Hurel2019,Yin2011,Zaman1996}, without explicitly addressing closed-loop
stabilization, constraint handling, or disturbance rejection.
Similarly, existing NMPC studies for flexible structures do not generally assess the
combined effects of impulsive disturbances, oscillatory actuator perturbations, actuator
saturation, and plant-model mismatch in a vehicle--on--cable setting~\cite{Artola2021}.
{As a result, closed-loop nonlinear predictive control of a balancing vehicle on a
flexible cable remains insufficiently explored, particularly when traversal accuracy,
roll--pitch stabilization, cable-vibration attenuation, and constraint satisfaction must be
achieved simultaneously.}
Related nonlinear modeling-and-control studies for underactuated balancing systems, such as
ball--beam and ball--pendulum configurations, further indicate the importance of
energy-based modeling and closed-loop validation for unstable mechanical systems
\cite{Tamimi2026BallPendulum}.

{Despite these advances, the existing literature does not provide a unified framework
that simultaneously combines: (i) a nonlinear finite-element/modal cable model retaining
both vertical and lateral transverse dynamics for a vehicle constrained to a tensioned
cable; (ii) nonlinear roll--pitch balance dynamics of an electric unicycle coupled to the
moving cable contact; (iii) a full nonlinear constrained receding-horizon NMPC formulation
posed directly on the reduced-order vehicle--cable dynamics; (iv) a frozen spatial
contact-point interpolation strategy applied across the NMPC prediction horizon; and
(v) closed-loop evaluation under disturbances, actuator limits, constraint tightening, and
plant-model mismatch.}

{The present work addresses this gap by developing a control-oriented nonlinear
modeling-and-NMPC framework for an electric unicycle traversing a tensioned cable. The
proposed frozen spatial interpolation strategy reduces repeated finite-element contact
operator updates within the optimizer while retaining nonlinear state evolution over the
prediction horizon.}

To address the identified research gap, this manuscript makes the following contributions:
\begin{enumerate}
\item A nonlinear reduced-order vehicle--cable interaction model is developed for an electric unicycle traversing a tensioned cable. The model couples the rigid-body roll and pitch dynamics of the vehicle with a nonlinear finite-element representation of the cable, retaining both vertical and lateral deformation modes through modal truncation.
\item A full nonlinear receding-horizon NMPC formulation is posed directly on the reduced-order dynamics, without resorting to linearization-based or LPV prediction models. The resulting nonlinear optimal control problem is solved online using IPOPT via CasADi, while explicitly enforcing actuator bounds, state constraints, and input-rate constraints.
\item {To reduce the computational burden of the nonlinear optimization problem,} a frozen spatial contact-point interpolation strategy is introduced for NMPC of the coupled vehicle--cable system.
\item Closed-loop robustness is systematically evaluated by separating nominal prediction and disturbed plant dynamics. External disturbances and modeling mismatch are injected exclusively into the nonlinear plant simulation, enabling direct assessment of disturbance rejection and robustness under realistic operating conditions.
\item {A practical numerical NMPC implementation is demonstrated, incorporating warm-starting, solver recovery through cold restarts, and a fallback control mechanism to maintain closed-loop simulation continuity in the presence of nonlinear solver failures.}
\end{enumerate}


The remainder of this manuscript is organized as follows.
Section~\ref{sec:model_cable} derives the coupled cable--unicycle dynamics, including the finite-element
cable formulation, the unicycle rigid-body model, and the resulting Lagrange equations of motion.
Section~\ref{sec:nonlinear_statespace_ocp} reformulates the coupled dynamics into a compact nonlinear
state-space form and introduces the modal reduction used for control-oriented prediction.
Section~\ref{sec:nocp} formulates the full nonlinear receding-horizon NMPC problem and presents the
proposed frozen spatial contact-point interpolation strategy for {improving the computational
tractability of the nonlinear optimization problem.}
Section~\ref{sec:algorithmic_implementation} summarizes the resulting closed-loop NMPC procedure and
its numerical implementation details.
Section~\ref{sec:results_cable_nmpc} reports numerical case studies and closed-loop results under
nominal and disturbed operating conditions, including quantitative tracking and robustness metrics.
Finally, Section~\ref{sec:conclusion} concludes the manuscript and outlines directions for future work.

\section{Coupled Cable--Unicycle Dynamics}
\label{sec:model_cable}
\begin{figure}[H]
\centering
\includegraphics[width=0.85\linewidth]{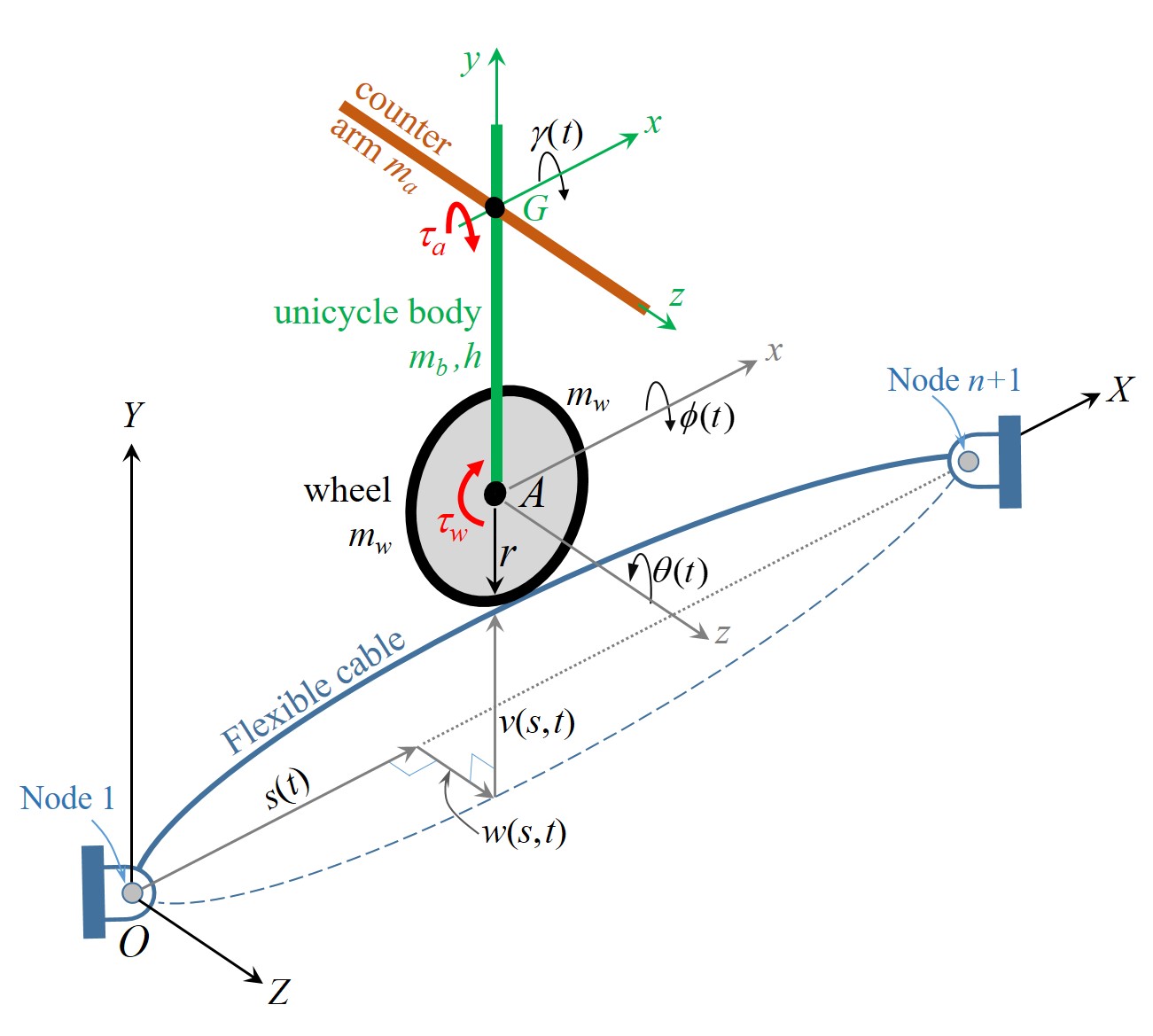}
\caption{Electric unicycle on a pretensioned flexible cable with moving contact $s(t)$ and transverse deflections $v(x,t)$ and $w(x,t)$.}
\label{fig:UConCable}
\end{figure}

This section presents the mathematical formulation of the flexible cable subsystem in the coupled
electric unicycle--cable system illustrated in Fig.~\ref{fig:UConCable}. The modeling framework is
developed to be consistent with finite-element discretization followed by model reduction,
facilitating subsequent integration with nonlinear model predictive control.

\begin{figure}[H]
\centering
\includegraphics[width=0.5\linewidth]{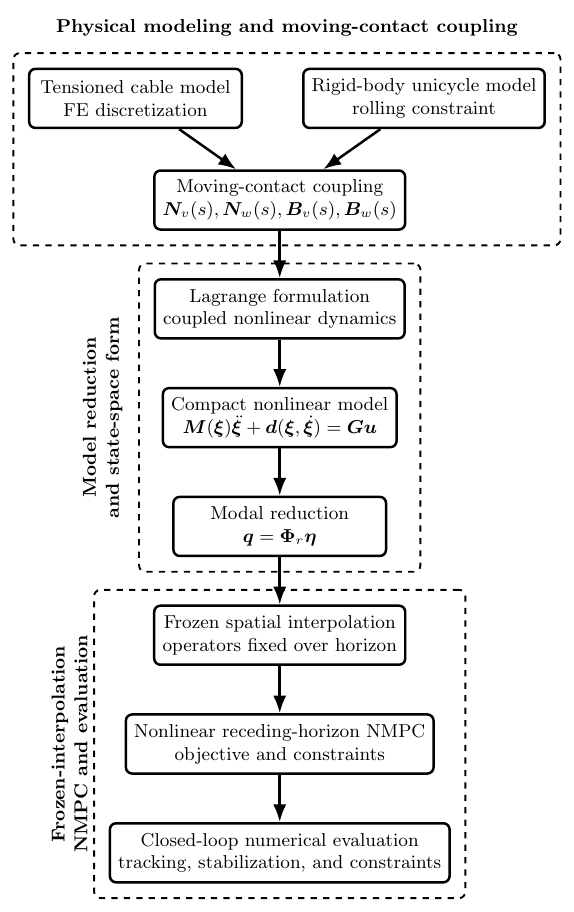}
\caption{{Modeling and control-oriented reduction framework for the electric unicycle on a tensioned cable.}}
\label{fig:modeling_control_framework}
\end{figure}

{Figure~\ref{fig:modeling_control_framework} summarizes the modeling procedure followed in this section and its connection to the subsequent NMPC formulation. The cable subsystem is first formulated using a tension-dominated finite-element model, while the unicycle is represented by axial, roll, pitch, and counter-arm coordinates. The two subsystems are coupled through moving-contact interpolation operators evaluated at the wheel--cable contact point. The resulting Lagrange-derived nonlinear model is then reduced using modal truncation and rewritten in compact nonlinear state-space form for use in the frozen-interpolation NMPC problem.}

The dynamic behavior of the supporting cable is first described at the continuous level.
Under the assumption of small-amplitude transverse vibrations and constant axial pretension,
the cable dynamics are governed by tension-dominated partial differential equations in the
vertical and lateral directions \cite{Li2014}.

The supporting cable has total length $L$, uniform mass per unit length $\rho A$, and is subjected to
a constant axial pretension $\Tt$. Small-amplitude transverse vibrations are assumed, such that
geometric nonlinearities and bending stiffness effects can be neglected. A Cartesian coordinate
system is introduced with the $x$-axis aligned with the undeformed cable centerline, the $y$-axis
denoting the vertical direction, and the $z$-axis denoting the lateral direction.

Let $v(x,t)$ and $w(x,t)$ denote the vertical and lateral displacements of the cable, respectively.
Here and in the sequel, evaluation of the cable displacement at the moving wheel--cable contact
point $x=s(t)$ is denoted by $v(s,t)$ and $w(s,t)$.

Under the above assumptions, the transverse dynamics of the taut cable are governed by the classical
wave equations
\begin{align}
\rho A \frac{\partial^2 v(x,t)}{\partial t^2}
- \Tt \frac{\partial^2 v(x,t)}{\partial x^2}
&= 0,
\label{eq:cable_pde_v}\\
\rho A \frac{\partial^2 w(x,t)}{\partial t^2}
- \Tt \frac{\partial^2 w(x,t)}{\partial x^2}
&= 0,
\label{eq:cable_pde_w}
\end{align}
which describe the vertical and lateral motion of a tension-dominated flexible cable \cite{JimenezOctavio2015}.

To enable numerical simulation, model reduction, and coupling with the rigid-body unicycle dynamics, the continuous cable equations are discretized in space using the finite-element method, with the cable displacements approximated over uniform elements with nodal degrees of freedom \cite{Karoumi1998,Sun2011}.

The cable is discretized into $n$ equal finite elements of length $l=L/n$, resulting in $n+1$ nodes.
Each node $i$ possesses two translational degrees of freedom: the vertical displacement $v_i(t)$ and
the lateral displacement $w_i(t)$. Accordingly, the local nodal displacement vector of element $e$
connecting nodes $i$ and $i+1$ is defined as
\begin{equation}
\bm q_e(t)=
\begin{bmatrix}
v_i & w_i & v_{i+1} & w_{i+1}
\end{bmatrix}^{\mathsf T}
\in \R^{4}.
\label{eq:qe_def}
\end{equation}

Within each element, the displacement fields are interpolated using linear shape functions \cite{Karoumi1998,Sun2011} defined
over the normalized local coordinate $\xi=x/l\in[0,1]$. The vertical and lateral displacements are
approximated as
\begin{align}
v(x,t)&=(1-\xi)v_i+\xi v_{i+1},
\label{eq:v_interp}\\
w(x,t)&=(1-\xi)w_i+\xi w_{i+1}.
\label{eq:w_interp}
\end{align}

These expressions can be written compactly in terms of the element nodal displacement vector as
\begin{equation}
v(x,t)=\bm N_v^e(x)\bm q_e(t), \qquad
w(x,t)=\bm N_w^e(x)\bm q_e(t),
\label{eq:disp_interp_compact}
\end{equation}
where the interpolation vectors for the vertical and lateral directions are given by
\begin{equation}
\bm N_v^e(x)=
\begin{bmatrix}
1-\xi & 0 & \xi & 0
\end{bmatrix},
\qquad
\bm N_w^e(x)=
\begin{bmatrix}
0 & 1-\xi & 0 & \xi
\end{bmatrix}.
\label{eq:shape_functions}
\end{equation}

The corresponding vertical and lateral slopes within the element are obtained by differentiating the
displacement fields with respect to the spatial coordinate $x$, yielding
\begin{equation}
\theta_v(x,t)=\bm B_v^e(x)\bm q_e(t), \qquad
\theta_w(x,t)=\bm B_w^e(x)\bm q_e(t),
\label{eq:slope_interp}
\end{equation}
with the slope interpolation vectors defined as
\begin{equation}
\bm B_v^e(x)
=\frac{\partial \bm N_v^e}{\partial x}
=
\begin{bmatrix}
-\dfrac{1}{l} & 0 & \dfrac{1}{l} & 0
\end{bmatrix},
\label{eq:Bv_def}
\end{equation}
\begin{equation}
\bm B_w^e(x)
=\frac{\partial \bm N_w^e}{\partial x}
=
\begin{bmatrix}
0 & -\dfrac{1}{l} & 0 & \dfrac{1}{l}
\end{bmatrix}.
\label{eq:Bw_def}
\end{equation}

Note that, due to the linear interpolation, the slopes $\theta_v(\cdot)$ and $\theta_w(\cdot)$
remain constant within each finite element. Here, $\theta_v(\cdot)$ and $\theta_w(\cdot)$ denote
cable slopes (spatial derivatives) and should not be confused with the unicycle pitch angle
$\theta(t)$ introduced later.

Using the principle of virtual work or standard finite-element procedures, the consistent element
mass and stiffness matrices associated with the tension-dominated cable are obtained as
\begin{equation}
\bm M_e
= \rho A l
\int_{0}^{1}
\left(
\bm N_v^{\mathsf T}\bm N_v
+
\bm N_w^{\mathsf T}\bm N_w
\right)
\,\dd\xi
=
\frac{\rho A l}{6}
\begin{bmatrix}
2 & 0 & 1 & 0\\
0 & 2 & 0 & 1\\
1 & 0 & 2 & 0\\
0 & 1 & 0 & 2
\end{bmatrix}
\in \R^{4\times4},
\label{eq:Me}
\end{equation}
\begin{equation}
\bm K_e
= \Tt l
\int_{0}^{1}
\left(
\bm B_v^{\mathsf T}\bm B_v
+
\bm B_w^{\mathsf T}\bm B_w
\right)
\,\dd\xi
=
\frac{\Tt}{l}
\begin{bmatrix}
1 & 0 & -1 & 0\\
0 & 1 & 0 & -1\\
-1 & 0 & 1 & 0\\
0 & -1 & 0 & 1
\end{bmatrix}
\in \R^{4\times4}.
\label{eq:Ke}
\end{equation}

After assembly of the element matrices and enforcement of the support boundary
conditions, the kinetic and potential energies of the discretized cable are
formulated to enable coupling with the unicycle dynamics through Lagrange's
equations.

The global degree-of-freedom vector of the discretized cable is defined as
\begin{equation}
\bm q_G(t)=
\begin{bmatrix}
v_1 & w_1 & v_2 & w_2 & \cdots & v_{n+1} & w_{n+1}
\end{bmatrix}^{\mathsf T}
\in \R^{2n+2}.
\label{eq:qG_def}
\end{equation}

Simply supported boundary conditions are imposed at both ends of the cable,
\begin{equation}
v_1=w_1=v_{n+1}=w_{n+1}=0,
\label{eq:bc}
\end{equation}
which eliminate the boundary degrees of freedom and yield the reduced global displacement vector
\begin{equation}
\bm q(t)=
\begin{bmatrix}
v_2 & w_2 & v_3 & w_3 & \cdots & v_n & w_n
\end{bmatrix}^{\mathsf T}
\in \R^{2n-2}.
\label{eq:q_reduced}
\end{equation}

The global mass and stiffness matrices, denoted by
$\bm M_c\in\R^{(2n-2)\times(2n-2)}$ and
$\bm K_c\in\R^{(2n-2)\times(2n-2)}$, are obtained by standard finite-element assembly of the element
matrices~\eqref{eq:Me}--\eqref{eq:Ke}, followed by enforcement of the boundary conditions
\eqref{eq:bc}.

The kinetic and potential energies of the discretized cable are finally expressed as
\begin{equation}
T_c=\frac{1}{2}\dot{\bm q}^{\mathsf T}\bm M_c\dot{\bm q},\qquad
V_c=\frac{1}{2}\bm q^{\mathsf T}\bm K_c\bm q,
\label{eq:cable_energies}
\end{equation}
where $\dot{\bm q}$ denotes the vector of nodal velocities.

Having established a finite-dimensional energetic description of the cable, the rigid-body dynamics
of the unicycle are now formulated and coupled to the cable through the moving contact point \cite{Abhishek2025}.

\subsection*{Unicycle rigid-body dynamics and energies}
\label{subsec:unicycle_dynamics_energies}

Having established a finite-dimensional energetic description of the cable subsystem, the
dynamics of the unicycle are now formulated and coupled to the cable through the moving
wheel--cable contact. The derivation follows a rigid-body modeling approach consistent with
the reference frames and kinematic definitions shown in Fig.~\ref{fig:UConCable}.

With these reference frames defined, the configuration of the unicycle is parameterized using
a minimal set of generalized coordinates that capture both its translational motion along the
cable and its rotational degrees of freedom.

We assume that the unicycle rolls without slipping along the cable. The inertial reference frame
$XYZ$ is attached to the left cable support, with the $Y$-axis oriented vertically upward and the
$X$-axis aligned with the cable, pointing toward the right support.

To describe the motion of each rigid component, a body-fixed reference frame $xyz$ is attached
to the unicycle body, counter arm, and wheel, with its origin located at the center of mass of
the corresponding component. The axes are defined such that the $x$-axis coincides with the
roll axis, the $y$-axis is initially vertical, and the $z$-axis corresponds to the pitch axis, as
illustrated in Fig.~\ref{fig:UConCable}.

Within this framework, the generalized coordinates of the unicycle are chosen as $s(t)$, the
axial position of the wheel--cable contact point measured from the left support; $\phi(t)$, the
roll angle describing rotation about the $x$-axis; $\theta(t)$, the pitch angle describing
rotation about the $z$-axis; and $\gamma(t)$, the angular position of the counter arm relative
to the body, measured about its $x$-axis.

\begin{figure}[H]
\centering
\includegraphics[width=0.3\linewidth]{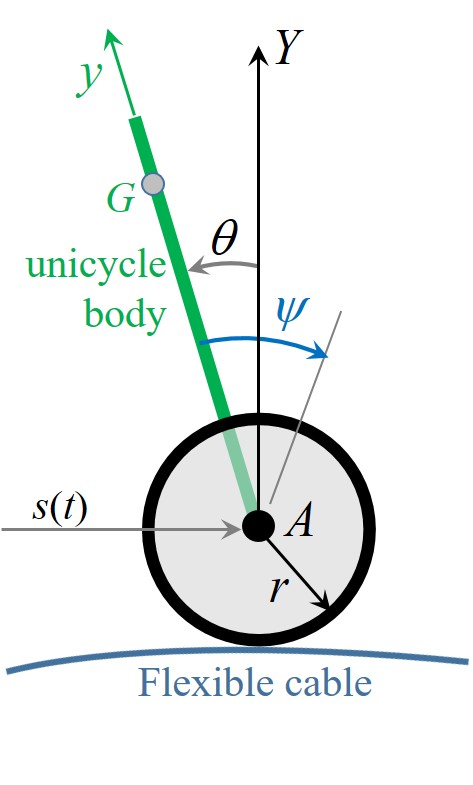}
\caption{Unicycle kinematics at the wheel–cable contact with pitch angle $\theta(t)$ and rolling constraint $s=r(\psi-\theta)$.}
\label{fig:Unicycle}
\end{figure}

The translational and rotational motions of the wheel are not independent but are constrained
by the rolling contact between the wheel and the cable. With reference to
Fig.~\ref{fig:Unicycle}, the axial position of the wheel is kinematically related to the wheel’s
angular displacement $\psi(t)$, defined as the clockwise rotation about the wheel $z$-axis
relative to the body, through the no-slip rolling constraint
\begin{equation}
s=r(\psi-\theta),
\label{eq:rolling_constraint}
\end{equation}
where $r$ denotes the wheel radius.

To describe the orientation of the unicycle body with respect to the inertial frame, a rotation
matrix of the body frame is constructed using the following rotation sequence:
\begin{equation}
x,y,z \xrightarrow{\bm R_x(\phi)} x',y',z' \xrightarrow{\bm R_z(\theta)} x'',y'',z''.
\label{eq:rotation_sequence}
\end{equation}

Roll about $x$ (unit vector, $\hat{x}$)
\begin{equation}
\bm R_x(\phi)=
\begin{bmatrix}
1 & 0 & 0 \\
0 & \cos\phi & -\sin\phi \\
0 & \sin\phi & \cos\phi
\end{bmatrix}
\label{eq:Rx}
\end{equation}
resulting in $x',y',z'$.

Pitch about $z'$ (unit vector, $\hat{z}'$)
\begin{equation}
\bm R_z(\theta)=
\begin{bmatrix}
\cos\theta & -\sin\theta & 0 \\
\sin\theta & \cos\theta & 0 \\
0 & 0 & 1
\end{bmatrix}
\label{eq:Rz}
\end{equation}
resulting in $x'',y'',z''$.

Carrying out the multiplication yields the rotation matrix that maps the final body-fixed frame
$x''y''z''$ into the inertial frame $XYZ$:
\begin{equation}
\bm R(\phi,\theta)=\bm R_x(\phi)\bm R_z(\theta)=
\begin{bmatrix}
\cos\theta & -\sin\theta & 0 \\
\cos\phi\sin\theta & \cos\phi\cos\theta & -\sin\phi \\
\sin\phi\sin\theta & \sin\phi\cos\theta & \cos\phi
\end{bmatrix}.
\label{eq:R_full}
\end{equation}

Based on the above orientation description, the absolute positions of the wheel center and the
unicycle body center of mass can now be expressed in the inertial frame. The position of the
wheel center, $A$, can be expressed as
\begin{equation}\label{eq:rA}
\bm r_A =
\begin{bmatrix}
s(t) \\
v(s,t)+r\cos\phi \\
w(s,t)+r\sin\phi
\end{bmatrix}.
\end{equation}

The position of the center of mass of the body, $G$, with respect to the wheel center, expressed
in the final body-fixed frame, is
\begin{equation}
\bm r''_{G/A}=h\,\hat{y}''=
h\begin{bmatrix}0\\1\\0\end{bmatrix},
\label{eq:rG_rel_doubleprime}
\end{equation}
where $h$ represents the distance between the wheel center $A$ and the body center of mass $G$.
Expressed in the inertial frame, this relative position becomes
\begin{equation}
\bm r_{G/A}=\bm R(\phi,\theta)\bm r''_{G/A}
=h\begin{bmatrix}
-\sin\theta\\
\cos\phi\cos\theta\\
\sin\phi\cos\theta
\end{bmatrix}.
\label{eq:rG_rel_inertial}
\end{equation}

Thus, the position of the body center of mass relative to the origin of the inertial frame is
\begin{equation}\label{eq:rG}
\bm r_G=\bm r_A+\bm r_{G/A}=
\begin{bmatrix}
s(t)-h\sin\theta\\
v(s,t)+r\cos\phi+h\cos\phi\cos\theta\\
w(s,t)+r\sin\phi+h\sin\phi\cos\theta
\end{bmatrix}.
\end{equation}

Since the wheel--cable contact point moves continuously along the cable span, the cable
displacements enter the unicycle kinematics through spatial interpolation of the reduced cable
coordinates \cite{app14073055}. To express the contact-point vertical and lateral displacements directly in terms
of the reduced global coordinates $\bm q(t)\in\mathbb{R}^{2n-2}$, global interpolation rows
$\bm N_v(s)$ and $\bm N_w(s)$ are introduced such that
\begin{equation}
v(s,t)=\bm N_v(s)\bm q(t), \qquad
w(s,t)=\bm N_w(s)\bm q(t).
\label{eq:vw_contact_interp}
\end{equation}

A compact representation over $s\in[0,L]$ can be formed using the Heaviside step function,
yielding
\begin{equation}
\bm N_v(s)=\sum_{i=1}^{n}
\Big[H\!\left(s-(i-1)l\right)-H\!\left(s-il\right)\Big]\bm N_{v}^{(i)}(s),
\label{eq:Nv_global}
\end{equation}
\begin{equation}
\bm N_w(s)=\sum_{i=1}^{n}
\Big[H\!\left(s-(i-1)l\right)-H\!\left(s-il\right)\Big]\bm N_{w}^{(i)}(s).
\label{eq:Nw_global}
\end{equation}

Similarly, the slopes of the cable at the moving contact point can be expressed over the entire
domain of $s$ as
\begin{equation}
\theta_v(s,t)=\bm B_v(s)\bm q(t), \qquad
\bm B_v(s)=\sum_{i=1}^{n}\Big[\big(H(s-(i-1)l)-H(s-il)\big)\bm B_v^{(i)}(s)\Big],
\label{eq:thetav_Bv_global}
\end{equation}
\begin{equation}
\theta_w(s,t)=\bm B_w(s)\bm q(t), \qquad
\bm B_w(s)=\sum_{i=1}^{n}\Big[\big(H(s-(i-1)l)-H(s-il)\big)\bm B_w^{(i)}(s)\Big].
\label{eq:thetaw_Bw_global}
\end{equation}

Substituting these interpolation expressions into
Eqs.~\eqref{eq:rA} and~\eqref{eq:rG} yields the following compact forms for the wheel center and
body center of mass positions:
\begin{equation}
\bm r_A =
\begin{bmatrix}
s(t)\\
\bm N_v(s)\bm q(t)+r\cos\phi\\
\bm N_w(s)\bm q(t)+r\sin\phi
\end{bmatrix},
\label{eq:rA_compact}
\end{equation}
\begin{equation}
\bm r_G =
\begin{bmatrix}
s(t)-h\sin\theta\\
\bm N_v(s)\bm q(t)+r\cos\phi+h\cos\phi\cos\theta\\
\bm N_w(s)\bm q(t)+r\sin\phi+h\sin\phi\cos\theta
\end{bmatrix}.
\label{eq:rG_compact}
\end{equation}

Differentiating the position vectors with respect to time yields the corresponding linear
velocities of the wheel center and the body center of mass. The velocity of the wheel center is
\begin{equation}
\bm v_A=\dot{\bm r}_A=
\begin{bmatrix}
\dot{s}\\
\bm B_v(s)\bm q(t)\,\dot{s}+\bm N_v(s)\dot{\bm q}(t)-r\dot{\phi}\sin\phi\\
\bm B_w(s)\bm q(t)\,\dot{s}+\bm N_w(s)\dot{\bm q}(t)+r\dot{\phi}\cos\phi
\end{bmatrix}.
\label{eq:vA}
\end{equation}

The velocity of the body center of mass is
\begin{equation}
\bm v_G=\dot{\bm r}_G=
\begin{bmatrix}
\dot{s}-h\dot{\theta}\cos\theta\\
\bm B_v(s)\bm q(t)\,\dot{s}+\bm N_v(s)\dot{\bm q}(t)
-r\dot{\phi}\sin\phi-h\dot{\phi}\sin\phi\cos\theta-h\dot{\theta}\cos\phi\sin\theta\\
\bm B_w(s)\bm q(t)\,\dot{s}+\bm N_w(s)\dot{\bm q}(t)
+r\dot{\phi}\cos\phi+h\dot{\phi}\cos\phi\cos\theta-h\dot{\theta}\sin\phi\sin\theta
\end{bmatrix}.
\label{eq:vG}
\end{equation}

In addition to translational motion, the rotational motion of each rigid component is described
through its angular velocity, which is defined below for the unicycle body, counter arm, and
wheel.

\paragraph{Angular velocities.}

Body:
\begin{equation}
\bm\omega_b=\dot{\phi}\,\hat{x}+\dot{\theta}\,\hat{z}'.
\label{eq:omega_b}
\end{equation}

Counter arm:
\begin{equation}
\bm\omega_a=(\dot{\phi}+\dot{\gamma})\,\hat{x}+\dot{\theta}\,\hat{z}'.
\label{eq:omega_a}
\end{equation}

Wheel:
\begin{equation}
\bm\omega_w=\dot{\phi}\,\hat{x}+(\dot{\psi}-\dot{\theta})\,\hat{z}'
=\dot{\phi}\,\hat{x}-\frac{\dot{s}}{r}\,\hat{z}'.
\label{eq:omega_w}
\end{equation}

For subsequent energy calculations, it is convenient to express these angular velocities in the
final body-fixed (double-primed) reference frame.

Angular velocities expressed in the final body-fixed frame (double-primed frame):

Body:
\begin{equation}
\bm\omega_b'' =
\dot{\phi}\,\bm R_z^{\mathsf T}(\theta)
\begin{bmatrix}
1\\
0\\
0
\end{bmatrix}
+\dot{\theta}
\begin{bmatrix}
0\\
0\\
1
\end{bmatrix}
=
\begin{bmatrix}
\dot{\phi}\cos\theta\\
-\dot{\phi}\sin\theta\\
\dot{\theta}
\end{bmatrix}.
\label{eq:omega_b_doubleprime}
\end{equation}

Counter arm:
\begin{equation}
\bm\omega_a'' =
(\dot{\phi}+\dot{\gamma})\,\bm R_z^{\mathsf T}(\theta)
\begin{bmatrix}
1\\
0\\
0
\end{bmatrix}
+\dot{\theta}
\begin{bmatrix}
0\\
0\\
1
\end{bmatrix}
=
\begin{bmatrix}
(\dot{\phi}+\dot{\gamma})\cos\theta\\
-(\dot{\phi}+\dot{\gamma})\sin\theta\\
\dot{\theta}
\end{bmatrix}.
\label{eq:omega_a_doubleprime}
\end{equation}

Wheel:
\begin{equation}
\bm\omega_w'' =
\dot{\phi}\,\bm R_z^{\mathsf T}(\theta)
\begin{bmatrix}
1\\
0\\
0
\end{bmatrix}
-\frac{\dot{s}}{r}
\begin{bmatrix}
0\\
0\\
1
\end{bmatrix}
=
\begin{bmatrix}
\dot{\phi}\cos\theta\\
-\dot{\phi}\sin\theta\\
-\dot{s}/r
\end{bmatrix}.
\label{eq:omega_w_doubleprime}
\end{equation}

With the translational and rotational kinematics fully defined, the kinetic and potential energies
of the unicycle components can now be formulated.

\paragraph{Unicycle energies.}

Body:
\begin{equation}
T_b=\frac{1}{2}m_b\,{\bm v}_G^{\mathsf T}{\bm v}_G
+\frac{1}{2}\bm\omega_b''^{\mathsf T}\bm I_b''\bm\omega_b''.
\label{eq:Tb}
\end{equation}

Counter arm:
\begin{equation}
T_a=\frac{1}{2}m_a\,{\bm v}_G^{\mathsf T}{\bm v}_G
+\frac{1}{2}\bm\omega_a''^{\mathsf T}\bm I_a''\bm\omega_a''.
\label{eq:Ta}
\end{equation}

Wheel:
\begin{equation}
T_w=\frac{1}{2}m_w\,{\bm v}_A^{\mathsf T}{\bm v}_A
+\frac{1}{2}\bm\omega_w''^{\mathsf T}\bm I_w''\bm\omega_w''.
\label{eq:Tw}
\end{equation}

Here, $m_b$, $m_w$ and $m_a$ denote the mass of the body, wheel and arm respectively.

$\bm I_b''$, $\bm I_w''$ and $\bm I_a''$ denote the inertia matrices of the body, wheel, and arm,
respectively, expressed in the double-primed reference frame. For simplicity, it is assumed that
the mass distribution of each component is symmetric with respect to the coordinate planes
passing through its center of mass. Under this assumption, the products of inertia vanish,
yielding diagonal inertia matrices:
\begin{equation}
\bm I_b''=
\begin{bmatrix}
I_{bx}'' & 0 & 0\\
0 & I_{by}'' & 0\\
0 & 0 & I_{bz}''
\end{bmatrix},\quad
\bm I_w''=
\begin{bmatrix}
I_{wx}'' & 0 & 0\\
0 & I_{wy}'' & 0\\
0 & 0 & I_{wz}''
\end{bmatrix},\quad
\bm I_a''=
\begin{bmatrix}
I_{ax}'' & 0 & 0\\
0 & I_{ay}'' & 0\\
0 & 0 & I_{az}''
\end{bmatrix}.
\label{eq:inertia_diagonal}
\end{equation}

The gravitational potential energy of each component is determined by its vertical position,
which depends explicitly on the cable deformation at the moving contact point.

Potential energy of the wheel:
\begin{equation}
V_w=m_w g\left(\bm N_v(s)\bm q(t)+r\cos\phi\right).
\label{eq:Vw}
\end{equation}

Potential energy of the body:
\begin{equation}
V_b=m_b g\left(\bm N_v(s)\bm q(t)+r\cos\phi+h\cos\phi\cos\theta\right).
\label{eq:Vb}
\end{equation}

Potential energy of the counter arm:
\begin{equation}
V_a=m_a g\left(\bm N_v(s)\bm q(t)+r\cos\phi+h\cos\phi\cos\theta\right),
\label{eq:Va}
\end{equation}
under the assumed simplified geometry, in which the body and counter arm share
the same center-of-mass location.

Combining the cable and unicycle energy expressions enables the construction of a unified
Lagrangian for the coupled system.

Total Lagrangian of the system:
\begin{equation}
\mathcal{L}=T_c+T_w+T_b+T_a-(V_c+V_w+V_b+V_a).
\label{eq:lagrangian_total}
\end{equation}

The equations of motion of the coupled cable--unicycle system are obtained by applying
Lagrange’s equations to the generalized coordinates introduced above. For clarity, the
resulting dynamics are grouped according to the cable degrees of freedom and the unicycle
degrees of freedom. {The expanded coordinate-wise equations corresponding to the cable and unicycle
degrees of freedom are provided in Appendix~\ref{app:detailed_model}.}

\textbf{Lagrange equations}

The generalized coordinates of the cable--unicycle system are therefore defined as
\begin{equation}
\begin{bmatrix}
\bm q & s & \phi & \theta & \gamma
\end{bmatrix}^{\mathsf T}\in\mathbb{R}^{2n+2}.
\label{eq:generalized_coordinates}
\end{equation}

\textbf{(1) Cable DOFs}

Applying Lagrange’s equations to the cable generalized coordinates yields the following set
of coupled nonlinear equations, which account for inertia, geometric coupling with the
unicycle motion, elastic restoring forces, and damping:
\begin{equation}
\frac{\dd}{\dd t}
\left(\frac{\partial\mathcal{L}}{\partial\dot{\bm q}}\right)
-\frac{\partial\mathcal{L}}{\partial\bm q}
+\bm C_c\dot{\bm q}
=\bm Q_q .
\label{eq:lagrange_cable_general}
\end{equation}

{The corresponding expanded cable-coordinate equation is reported in
Appendix~\ref{app:detailed_model}. In that expression,}
$m_u=(m_w+m_b+m_a)$ represents the total mass of the unicycle, $m_{ab}=(m_a+m_b)$ and
$\bm C_c=\alpha\bm M_c+\beta\bm K_c\in\mathbb{R}^{(2n-2)\times(2n-2)}$
represents Rayleigh damping in the cable \cite{Abhishek2025}.

\textbf{(2) Unicycle DOFs}

The equations governing the unicycle motion are obtained by applying Lagrange’s equations
to the remaining generalized coordinates associated with the axial position and rotational
degrees of freedom. {The expanded axial, roll, pitch, and counter-arm equations are
provided in Appendix~\ref{app:detailed_model}.}

\paragraph{Axial motion}

\begin{equation}
\frac{\dd}{\dd t}
\left(\frac{\partial\mathcal{L}}{\partial\dot{s}}\right)
-\frac{\partial\mathcal{L}}{\partial s}
=Q_s .
\label{eq:lagrange_s_general}
\end{equation}

\paragraph{Roll dynamics}

\begin{equation}
\frac{\dd}{\dd t}
\left(\frac{\partial\mathcal{L}}{\partial\dot{\phi}}\right)
-\frac{\partial\mathcal{L}}{\partial \phi}
=Q_{\phi}.
\label{eq:lagrange_phi_general}
\end{equation}

\paragraph{Pitch dynamics}

\begin{equation}
\frac{\dd}{\dd t}
\left(\frac{\partial\mathcal{L}}{\partial\dot{\theta}}\right)
-\frac{\partial\mathcal{L}}{\partial \theta}
=Q_{\theta}.
\label{eq:lagrange_theta_general}
\end{equation}

\paragraph{Counter-arm dynamics}

\begin{equation}
\frac{\dd}{\dd t}
\left(\frac{\partial\mathcal{L}}{\partial\dot{\gamma}}\right)
-\frac{\partial\mathcal{L}}{\partial \gamma}
=Q_{\gamma}.
\label{eq:lagrange_gamma_general}
\end{equation}

\begin{equation}
I_{ax}\ddot{\phi}+I_{ax}\ddot{\gamma}=Q_{\gamma}.
\label{eq:gamma_eom}
\end{equation}

The unicycle is controlled via two inputs: the torque $\tau_w(t)$ applied about the wheel local
$z$-axis to the rotational coordinate $\psi(t)$, and the torque $\tau_a(t)$ applied about the
counter-arm local $x$-axis to the rotational coordinate $\gamma(t)$. The net virtual work of these
inputs is
\begin{equation}
\delta W=\tau_w\,\delta\psi+\tau_a\,\delta\gamma,
\qquad
\psi=\frac{s}{r}+\theta .
\label{eq:virtual_work_inputs}
\end{equation}

From the virtual work expression, the generalized forces associated with each coordinate are
obtained as follows.

The resulting generalized forces are
\begin{equation}
\bm Q_q
=\tau_w\frac{\partial\psi}{\partial\bm q}
+\tau_a\frac{\partial\gamma}{\partial\bm q}
=\bm 0,
\label{eq:Qq}
\end{equation}
\begin{equation}
Q_s
=\tau_w\frac{\partial\psi}{\partial s}
+\tau_a\frac{\partial\gamma}{\partial s}
=\frac{\tau_w}{r},
\label{eq:Qs}
\end{equation}
\begin{equation}
Q_{\phi}
=\tau_w\frac{\partial\psi}{\partial\phi}
+\tau_a\frac{\partial\gamma}{\partial\phi}
=0,
\label{eq:Qphi}
\end{equation}
\begin{equation}
Q_{\theta}
=\tau_w\frac{\partial\psi}{\partial\theta}
+\tau_a\frac{\partial\gamma}{\partial\theta}
=\tau_w,
\label{eq:Qtheta}
\end{equation}
\begin{equation}
Q_{\gamma}
=\tau_w\frac{\partial\psi}{\partial\gamma}
+\tau_a\frac{\partial\gamma}{\partial\gamma}
=\tau_a.
\label{eq:Qgamma}
\end{equation}

\section{Nonlinear State-Space Model for Control}
\label{sec:nonlinear_statespace_ocp}

Building directly on the Lagrange-derived equations of motion presented in the previous
section, this section reformulates the coupled unicycle--cable dynamics into a compact
first-order nonlinear state-space representation suitable for simulation, analysis, and
control synthesis. The formulation preserves the full nonlinear structure of the model and
serves as the foundation for subsequent model reduction and nonlinear optimal control design.
{The expanded coordinate-wise equations and the detailed mass-matrix and drift-vector
entries are provided in Appendix~\ref{app:detailed_model}; the main text retains the
compact control-oriented representation used for the NMPC formulation.}

\subsection*{Coordinates, state vector, and inputs}
\label{subsec:state_definition}

Let the generalized coordinate vector be
\begin{equation}
\bm \xi(t)=
\begin{bmatrix}
\bm q^{\mathsf T}(t) & s(t) & \phi(t) & \theta(t) & \gamma(t)
\end{bmatrix}^{\mathsf T}\in\mathbb{R}^{2n+2},
\label{eq:xi_def}
\end{equation}
which collects the cable nodal coordinates together with the unicycle axial and rotational
degrees of freedom introduced earlier. Based on this definition, the full nonlinear state
vector and control inputs are defined as
\begin{equation}
\bm x(t)=
\begin{bmatrix}
\bm \xi^{\mathsf T}(t) & \dot{\bm \xi}^{\mathsf T}(t)
\end{bmatrix}^{\mathsf T}
=
\begin{bmatrix}
\bm q^{\mathsf T} & s & \phi & \theta & \gamma &
\dot{\bm q}^{\mathsf T} & \dot s & \dot\phi & \dot\theta & \dot\gamma
\end{bmatrix}^{\mathsf T}\in\mathbb{R}^{4n+4},
\label{eq:x_def}
\end{equation}
\begin{equation}
\bm u(t)=
\begin{bmatrix}
\tau_w(t)\\
\tau_a(t)
\end{bmatrix}.
\label{eq:u_def}
\end{equation}

As before, the reduced cable interpolation and slope rows evaluated at the moving contact
point $s(t)\in[0,L]$ are denoted by
$\bm N_v(s),\bm N_w(s),\bm B_v(s),\bm B_w(s)\in\mathbb{R}^{1\times(2n-2)}$.
For compactness of notation, define the slope-related scalars
\begin{equation}
\nu_v := \bm B_v(s)\bm q,\qquad
\nu_w := \bm B_w(s)\bm q,\qquad
\dot\nu_v := \bm B_v(s)\dot{\bm q},\qquad
\dot\nu_w := \bm B_w(s)\dot{\bm q}.
\label{eq:nu_defs}
\end{equation}

\subsection*{Compact second-order nonlinear form}
\label{subsec:compact_second_order}

For control-oriented representation and numerical implementation, the coupled
unicycle--cable dynamics are collected into the compact second-order form
\begin{equation}
\bm M(\bm \xi)\,\ddot{\bm \xi}+\bm d(\bm \xi,\dot{\bm \xi})=\bm G\,\bm u,
\label{eq:nonlinear_compact}
\end{equation}
where $\bm M(\bm \xi)\in\mathbb{R}^{(2n+2)\times(2n+2)}$ is the configuration-dependent mass
matrix, $\bm d(\bm \xi,\dot{\bm \xi})\in\mathbb{R}^{2n+2}$ collects all damping, stiffness, gravity,
and velocity-dependent nonlinear terms, and $\bm G\in\mathbb{R}^{(2n+2)\times 2}$ maps the
control torques into the generalized coordinates. {The detailed entries of
\(\bm M(\bm \xi)\) and \(\bm d(\bm \xi,\dot{\bm \xi})\) are listed in
Appendix~\ref{app:detailed_model}.}

\subsection*{Mass matrix structure}
\label{subsec:mass_matrix_structure}

To make the nonlinear dynamics explicit and suitable for implementation, the configuration-
dependent mass matrix is written in block-partitioned form. The mass matrix is then given by
\begin{equation}
\bm M(\bm \xi)=
\begin{bmatrix}
\bm M_{qq} & \bm M_{qs} & \bm M_{q\phi} & \bm M_{q\theta} & \bm 0\\
\bm M_{sq} & M_{ss} & M_{s\phi} & M_{s\theta} & 0\\
\bm M_{\phi q} & M_{\phi s} & M_{\phi\phi} & 0 & M_{\phi\gamma}\\
\bm M_{\theta q} & M_{\theta s} & 0 & M_{\theta\theta} & 0\\
\bm 0 & 0 & M_{\gamma\phi} & 0 & M_{\gamma\gamma}
\end{bmatrix},
\label{eq:M_block}
\end{equation}
{with the individual block entries reported in Appendix~\ref{app:detailed_model}.}
All unspecified blocks are identically zero.

\subsection*{Input mapping and nonlinear drift terms}
\label{subsec:input_and_drift}

The generalized forces induced by the wheel and counter-arm torques follow directly from the
virtual-work formulation established previously, yielding
\begin{equation}
\bm Q_q=\bm 0,\qquad
Q_s=\frac{\tau_w}{r},\qquad
Q_{\phi}=0,\qquad
Q_{\theta}=\tau_w,\qquad
Q_{\gamma}=\tau_a,
\label{eq:generalized_forces_compact}
\end{equation}
and the constant input mapping matrix
\begin{equation}
\bm G=
\begin{bmatrix}
\bm 0_{(2n-2)\times 2}\\
\frac{1}{r} & 0\\
0 & 0\\
1 & 0\\
0 & 1
\end{bmatrix}.
\label{eq:Gmatrix}
\end{equation}

The remaining nonlinear terms are collected in the drift vector
$\bm d(\bm \xi,\dot{\bm \xi})$, which is partitioned as
\begin{equation}
\bm d(\bm \xi,\dot{\bm \xi})=
\begin{bmatrix}
\bm d_q^{\mathsf T} &
h_s &
h_{\phi} &
h_{\theta} &
h_{\gamma}
\end{bmatrix}^{\mathsf T},
\label{eq:d_partition}
\end{equation}
{where the individual components are provided in Appendix~\ref{app:detailed_model}.}

\subsection*{First-order nonlinear state-space representation}
\label{subsec:first_order_ss}

To obtain a standard state-space representation suitable for nonlinear control and optimization,
the second-order dynamics are rewritten in first-order form.
Introducing the first-order state variables $\bm x_1=\bm \xi$ and $\bm x_2=\dot{\bm \xi}$,
the complete nonlinear state-space model is written as
\begin{equation}
\dot{\bm x}_1=\bm x_2,
\label{eq:ss1}
\end{equation}
\begin{equation}
\dot{\bm x}_2
=\bm M(\bm x_1)^{-1}\Big(\bm G\,\bm u-\bm d(\bm x_1,\bm x_2)\Big).
\label{eq:ss2}
\end{equation}
Equivalently,
\begin{equation}
{
\dot{\bm x}=
\begin{bmatrix}
\bm x_2\\
\bm M(\bm x_1)^{-1}\Big(\bm G\,\bm u-\bm d(\bm x_1,\bm x_2)\Big)
\end{bmatrix}.}
\label{eq:full_nonlinear_statespace}
\end{equation}

\subsection*{Modal reduction and nonlinear optimal control formulation}
\label{subsec:modal_reduction}

To obtain a control-oriented model and suppress high-frequency numerical dynamics, the cable
coordinates are expressed in a truncated modal basis. Consider the free-vibration eigenproblem
\begin{equation}
\bm K_c \bm\Phi = \bm M_c \bm\Phi \bm\Omega^2,
\label{eq:eig_problem}
\end{equation}
where the columns of $\bm\Phi$ are mass-normalized mode shapes and
$\bm\Omega=\mathrm{diag}(\omega_1,\omega_2,\dots)$ contains the natural frequencies.
Retaining the first $r$ dominant modes yields
\begin{equation}
\bm q(t)=\bm\Phi_r \bm\eta(t),\qquad
\dot{\bm q}(t)=\bm\Phi_r \dot{\bm\eta}(t),
\label{eq:modal_expansion}
\end{equation}
with the orthonormality conditions
\begin{equation}
\bm\Phi_r^{\mathsf T}\bm M_c \bm\Phi_r=\bm I_r,\qquad
\bm\Phi_r^{\mathsf T}\bm K_c \bm\Phi_r=\bm\Omega_r^2.
\label{eq:modal_orthonormality}
\end{equation}

Substituting the modal expansion into the nonlinear dynamics leads to a reduced-order model
that preserves the second-order structure and is used directly in the nonlinear optimal
control problem formulated next.

\section{Nonlinear optimal control problem}
\label{sec:nocp}

The coupled cable--unicycle system derived in the previous sections exhibits
strong nonlinearities, configuration-dependent inertia, and bidirectional
vehicle--structure interaction through a continuously moving contact point.
In addition, the system is subject to hard state and input constraints arising
from balance limits, actuator saturation, and the finite cable span.
These characteristics render linear control approaches insufficient for
robust traversal and motivate the use of nonlinear optimal control.
Algorithmic developments for nonlinear model predictive control targeting fast and
computationally demanding systems have been investigated with emphasis on efficient numerical
realization and {reduced computational burden} \cite{Tamimi2011VDI}.

In this work, the control objective is to steer the unicycle along the cable
from an initial contact position $s(0)=s_0$ to a desired terminal position
$s_f$, while maintaining roll and pitch stability and suppressing cable
vibrations.
This objective is achieved by solving a constrained nonlinear optimal control
problem (OCP) based on the modal-reduced nonlinear dynamics.

\subsubsection*{Modal-reduced nonlinear dynamics}

Following the modal reduction introduced in
Section~\ref{subsec:modal_reduction}, the generalized coordinates are
\begin{equation}
\bm y(t)=
\begin{bmatrix}
\bm\eta^{\mathsf T}(t) & s(t) & \phi(t) & \theta(t) & \gamma(t)
\end{bmatrix}^{\mathsf T}\in\mathbb{R}^{r+4},
\end{equation}
and the corresponding state vector is
\begin{equation}
\bm x(t)=
\begin{bmatrix}
\bm y^{\mathsf T}(t) & \dot{\bm y}^{\mathsf T}(t)
\end{bmatrix}^{\mathsf T}
=
\begin{bmatrix}
\bm\eta^{\mathsf T} & s & \phi & \theta & \gamma &
\dot{\bm\eta}^{\mathsf T} & \dot{s} & \dot{\phi} & \dot{\theta} & \dot{\gamma}
\end{bmatrix}^{\mathsf T}.
\end{equation}

The modal-reduced coupled dynamics preserve the second-order structure
\begin{equation}
\bar{\bm M}(\bm y)\,\ddot{\bm y}
+\bar{\bm d}(\bm y,\dot{\bm y})
=
\bar{\bm G}\,\bm u,
\label{eq:modal_second_order}
\end{equation}
where $\bar{\bm M}(\bm y)$ and $\bar{\bm d}(\bm y,\dot{\bm y})$ are obtained from
$\bm M(\bm\xi)$ and $\bm d(\bm\xi,\dot{\bm\xi})$ in
\eqref{eq:nonlinear_compact} by substituting
$\bm q=\bm\Phi_r\bm\eta$ and $\dot{\bm q}=\bm\Phi_r\dot{\bm\eta}$.
The input mapping $\bar{\bm G}$ is identical to $\bm G$ in
\eqref{eq:Gmatrix} with appropriate dimension reduction.

Introducing first-order state variables $\bm x_1=\bm y$ and
$\bm x_2=\dot{\bm y}$ yields the nonlinear state-space form
\begin{equation}
\dot{\bm x}_1=\bm x_2,
\qquad
\dot{\bm x}_2=
\bar{\bm M}(\bm x_1)^{-1}
\Big(\bar{\bm G}\bm u-\bar{\bm d}(\bm x_1,\bm x_2)\Big),
\label{eq:ss_modal_reduced}
\end{equation}
which is compactly written as $\dot{\bm x}=\bar{\bm f}(\bm x,\bm u)$.

\subsubsection*{Continuous-time nonlinear OCP}

Let $t_k$ denote the current sampling instant.
Over the finite horizon $[t_k,t_k+T]$, the nonlinear optimal control problem is
formulated as
\begin{equation}
\label{eq:ocp_continuous}
\begin{aligned}
\min_{\bm u(\cdot)}\quad
& J =
\int_{t_k}^{t_k+T}
\Big(
q_s (s-s_f)^2
+q_\phi \phi^2
+q_\theta \theta^2
+\dot{\bm\eta}^{\mathsf T}\bm Q_\eta\dot{\bm\eta}
+q_{\dot s}(\dot{s}-\dot{s}_{\mathrm{ref}})^2
+q_{\dot\phi}\dot{\phi}^2
+q_{\dot\theta}\dot{\theta}^2
+\bm u^{\mathsf T}\bm R_u\bm u
\Big)\,\dd t
\\
&\quad
+(\bm x(t_k+T)-\bm x_f)^{\mathsf T}
\bm P_f
(\bm x(t_k+T)-\bm x_f)
\\[0.5em]
\text{s.t.}\quad
& \dot{\bm x}(t)=\bar{\bm f}(\bm x(t),\bm u(t)),
\qquad
\bm x(t_k)=\bm x_k,
\\
& \tau_w^{\min}\le\tau_w(t)\le\tau_w^{\max},
\qquad
\tau_a^{\min}\le\tau_a(t)\le\tau_a^{\max},
\\
& 0\le s(t)\le L,
\\
& |\phi(t)|\le\phi_{\max},
\qquad
|\theta(t)|\le\theta_{\max}.
\end{aligned}
\end{equation}

Here $\dot{s}_{\mathrm{ref}}$ denotes a reference axial velocity chosen to
ensure progress toward $s_f$,
$\bm Q_\eta\succeq0$ penalizes modal vibration velocities,
$\bm R_u=\mathrm{diag}(R_w,R_a)\succ0$ penalizes control effort, and
$\bm P_f\succeq0$ is a terminal weighting matrix.
The matrix $\bm R_u$ should not be confused with the rotation matrix
$\bm R(\phi,\theta)$ defined in \eqref{eq:R_full}.
The terminal target $\bm x_f$ enforces $s(t_k+T)=s_f$ with zero angular and
modal velocities.

\subsubsection*{Discrete-time transcription}

For numerical solution, the continuous-time OCP
\eqref{eq:ocp_continuous} is transcribed into a discrete-time nonlinear
program using a sampling time $T_s$ and prediction horizon $N=T/T_s$.
The selection of the prediction horizon length and discretization intervals plays a critical role
in balancing numerical accuracy and computational efficiency in nonlinear MPC implementations
\cite{Tamimi2021TIMC}.

The discrete-time dynamics are obtained via fourth-order Runge--Kutta integration,
\begin{equation}
\bm x_{k+1}
=
\bm f_d(\bm x_k,\bm u_k),
\label{eq:rk4_discrete}
\end{equation}
where $\bm f_d(\cdot)$ represents the RK4 discretization of
$\dot{\bm x}=\bar{\bm f}(\bm x,\bm u)$.

The discrete-time OCP is then
\begin{equation}
\label{eq:ocp_discrete}
\begin{aligned}
\min_{\{\bm u_k\}_{k=0}^{N-1}}\quad
& \sum_{k=0}^{N-1}
\ell(\bm x_k,\bm u_k)
+\ell_f(\bm x_N)
\\
\text{s.t.}\quad
& \bm x_{k+1}=\bm f_d(\bm x_k,\bm u_k),
\qquad
\bm x_0=\bm x(t_k),
\\
& \bm u_k\in\mathcal U,\qquad
\bm x_k\in\mathcal X,
\end{aligned}
\end{equation}
where $\mathcal U$ and $\mathcal X$ collect the input and state constraints,
respectively.

\subsubsection*{Receding-horizon implementation}

The discrete-time OCP \eqref{eq:ocp_discrete} is embedded in a receding-horizon
control (RHC) strategy.
At each sampling instant, the OCP is solved using the current measured state,
only the first control input is applied to the plant, and the horizon is shifted
forward by one step.

This receding-horizon implementation naturally accommodates nonlinear dynamics,
hard constraints, and moving-contact effects, and forms the basis for the
nonlinear model predictive control (NMPC) simulations presented in the
following section.

While the formulation above defines the nonlinear NMPC problem, its direct
solution remains computationally demanding due to the moving contact-point
dependence addressed next.

\subsection{Frozen contact-point interpolation in nonlinear NMPC}
\label{subsec:frozen_interpolation}

A key computational challenge in applying nonlinear model predictive control
to the coupled cable--unicycle system arises from the moving wheel--cable
contact.
As shown in Section~\ref{sec:model_cable}, the cable deformation enters the
unicycle kinematics and dynamics through spatial interpolation and slope
operators evaluated at the instantaneous contact position $s(t)$, namely
$\bm N_v(s)$, $\bm N_w(s)$, $\bm B_v(s)$, and $\bm B_w(s)$.
These operators are piecewise-defined over the finite-element mesh and depend
explicitly on the axial coordinate, which itself is a state of the system.

In a standard NMPC formulation, the prediction model must be evaluated at every
stage of the prediction horizon using the predicted state trajectory.
For moving-contact systems, this implies repeated evaluation of the
finite-element interpolation operators at predicted contact positions
$s_{k+i}$.
As a result, the structure of the prediction model changes across the horizon,
leading to increased symbolic complexity, dense Jacobians and Hessians, and
significant computational overhead within the nonlinear programming solver.
This effect becomes particularly severe when combined with finite-element
discretization, modal reduction, and long prediction horizons, making direct NMPC implementations computationally demanding for online control.

To address this issue, a frozen contact-point interpolation strategy is adopted.
At each sampling instant $t_k$, the current contact position $s_k$ is measured
or estimated, and the reduced interpolation and slope operators are evaluated
once at this position,
\[
\bar{\bm N}_v=\bm N_v(s_k),\quad
\bar{\bm N}_w=\bm N_w(s_k),\quad
\bar{\bm B}_v=\bm B_v(s_k),\quad
\bar{\bm B}_w=\bm B_w(s_k).
\]
These operators are then held constant over the entire prediction horizon.
Only the spatial interpolation operators are frozen; all state variables,
including the axial position $s(t)$, the rotational angles, and the modal
coordinates, continue to evolve according to the full nonlinear dynamics over
the prediction horizon.

With the interpolation operators frozen, the prediction model used inside the
NMPC problem retains its full nonlinear structure while becoming parametrically
fixed over the horizon.
The resulting prediction dynamics can be written as
\[
\dot{\bm x}(t)
=
\bar{\bm f}\!\left(
\bm x(t),\bm u(t);
\bar{\bm N}_v,\bar{\bm N}_w,\bar{\bm B}_v,\bar{\bm B}_w
\right),
\]
where the contact coordinate $s(t)$ remains an explicit dynamic state.
This formulation yields a nonlinear optimal control problem with frozen
parameters, which can be solved efficiently using a standard interior-point
solver without repeated finite-element operator assembly.

The proposed strategy is related to frozen-parameter NMPC approaches, in which
slowly varying quantities are held constant over the prediction horizon to
reduce computational burden. Here, however, the frozen quantities are spatial
finite-element interpolation operators associated with a moving contact
constraint, rather than time-varying system matrices or scheduling parameters.
To the authors’ knowledge, this constitutes a novel application of frozen
prediction in nonlinear NMPC for vehicle--structure interaction systems.

Freezing the contact-point interpolation operators significantly reduces the
computational complexity of the NMPC problem by fixing the symbolic structure
of the prediction model, eliminating element-membership logic inside the
optimizer, and improving sparsity and solver robustness.
These properties improve the computational tractability of the finite-element-based NMPC formulation { and provide a basis for future investigation using dedicated real-time optimization implementations.}

The frozen interpolation strategy introduces an approximation by neglecting the
variation of the contact-point operators over a single prediction horizon.
This approximation is justified by the short sampling time, the limited contact
displacement per horizon, and the receding-horizon implementation.
At each sampling instant, the operators are re-evaluated using the updated
measured contact position, ensuring that any model mismatch is corrected in
closed loop.

This frozen contact-point interpolation strategy constitutes a core
methodological contribution of this work, as it enables the integration of
high-fidelity finite-element cable models with full nonlinear NMPC for
moving-contact systems, without resorting to LPV scheduling or repeated
finite-element operator assembly inside the optimizer.

\subsection{Stability and recursive-feasibility discussion}
\label{subsec:stability_recursive_feasibility}

{For constrained receding-horizon NMPC, nominal recursive feasibility and local closed-loop stability are commonly established by augmenting the finite-horizon problem with suitable terminal components. These components usually include a terminal penalty \(V_f(\bm x)\), a terminal admissible set \(\mathcal X_f\), and a local terminal feedback law \(\kappa_f(\bm x)\). For the discrete-time prediction model}
\begin{equation}
\bm x_{k+1}=\bm f_d(\bm x_k,\bm u_k),
\end{equation}
{the terminal set is required to be positively invariant under \(\kappa_f\), and the terminal penalty is required to satisfy the local decrease condition}
\begin{equation}
V_f\!\left(\bm f_d(\bm x,\kappa_f(\bm x))\right)-V_f(\bm x)
\leq
-\ell\!\left(\bm x,\kappa_f(\bm x)\right),
\qquad
\bm x\in\mathcal X_f .
\label{eq:terminal_decrease_condition}
\end{equation}
{When these conditions hold, the shifted optimal input sequence, appended with the terminal feedback action, provides a feasible candidate solution at the next sampling instant. In addition, the optimal finite-horizon value function \(V_N(\bm x_k)\) satisfies the standard decrease relation}
\begin{equation}
V_N(\bm x_{k+1})-V_N(\bm x_k)
\leq
-\ell(\bm x_k,\bm u_k^\star),
\label{eq:nmpc_value_decrease}
\end{equation}
{which provides the usual Lyapunov-based argument for nominal recursive feasibility and local closed-loop stability with respect to the terminal target or terminal set~\cite{Mayne2000NMPC,Chen1998NMPC,Rawlings2017,Grune2017NMPC}.}

{The NMPC formulation adopted in this study follows this constrained receding-horizon framework and includes a terminal penalty, actuator limits, input-rate constraints, balance-angle constraints, and cable-span constraints. The prediction model is the reduced-order nonlinear unicycle--cable model, which contains the balancing dynamics of the vehicle, modal cable coordinates, moving-contact coupling, and nonlinear state-dependent interaction terms. For this class of coupled vehicle--structure dynamics, the explicit computation of a certified terminal invariant set is challenging because the admissible set is simultaneously shaped by balance constraints, actuator saturation, rate limits, cable-span restrictions, and the modal coupling induced by the moving contact point.}

{Therefore, the terminal penalty is used as a stabilizing numerical component within the finite-horizon optimal-control problem. Since an explicit terminal invariant set and terminal feedback law are not constructed, the formulation is not accompanied by a formal recursive-feasibility or closed-loop stability proof. Instead, the numerical implementation is assessed through the closed-loop simulated trajectories, observed constraint satisfaction, solver behavior, and robustness metrics under the scenario studies presented later in Section~\ref{sec:results_cable_nmpc}. The construction of certified terminal sets and terminal feedback laws for the full nonlinear unicycle--cable model is identified as an important direction for future theoretical development.}

\section{Algorithmic implementation of frozen NMPC}
\label{sec:algorithmic_implementation}

{This section presents the closed-loop algorithmic realization of the nonlinear model predictive control (NMPC) scheme with frozen contact-point interpolation in a MATLAB/CasADi/IPOPT prototyping environment.}
The implementation directly follows the nonlinear optimal control formulation
in Section~\ref{sec:nocp} and the frozen-interpolation prediction model defined
in \eqref{eq:frozen_prediction_model}.

The controller is implemented in a receding-horizon fashion.
At each sampling instant $t_k$, the current system state $\bm x_k$ and the
wheel--cable contact position $s_k$ are measured or estimated.
Frozen contact-point interpolation is implemented as described in
Section~\ref{sec:nocp}.

This yields a parametrically fixed nonlinear model of the form
\eqref{eq:frozen_prediction_model}, while all state variables—including the
axial position, rotational coordinates, and modal states—continue to evolve
according to the full nonlinear dynamics.

Using the frozen prediction model, the continuous-time optimal control problem
defined in Section~\ref{sec:nocp} is transcribed into a discrete-time nonlinear
program via fourth-order Runge--Kutta integration.
The resulting finite-dimensional optimization problem is solved using a
standard interior-point method to obtain an optimal control sequence over the
prediction horizon \cite{Diehl2005, Rawlings2017, Tamimi2011VDI}.

As in conventional NMPC, only the first control input of the optimal sequence
is applied to the plant.
At the next sampling instant, the horizon is shifted forward, the contact-point
operators are re-evaluated using the updated measurements, and the optimization
is repeated.

From an algorithmic standpoint, freezing the contact-point interpolation
operators results in a fixed symbolic structure of the nonlinear program within
each NMPC iteration.
This property eliminates horizon-wise changes in the prediction model caused by
element membership and spatial interpolation, leading to improved solver
robustness and more predictable computational complexity.
Consequently, the online computational burden scales primarily with the state
dimension and prediction horizon, rather than with the spatial discretization
of the cable.

The complete closed-loop NMPC procedure with frozen contact-point interpolation
is summarized in Algorithm~\ref{alg:frozen_nmpc}.

\begin{algorithm}[htbp!]
\caption{Nonlinear NMPC with frozen contact-point interpolation}
\label{alg:frozen_nmpc}
\begin{algorithmic}[1]
\State \textbf{Input:} Sampling time $T_s$, horizon length $N$, modal basis
$\bm\Phi_r$, constraint sets $\mathcal X$, $\mathcal U$
\State \textbf{Initialize:} $k \gets 0$, measure or estimate initial state
$\bm x_0$

\While{control task not completed}

    \State Measure current state $\bm x_k$ and contact position $s_k$

    \State Evaluate reduced cable operators at $s_k$:
    \[
    \bar{\bm N}_v \gets \bm N_v(s_k),\quad
    \bar{\bm N}_w \gets \bm N_w(s_k),\quad
    \bar{\bm B}_v \gets \bm B_v(s_k),\quad
    \bar{\bm B}_w \gets \bm B_w(s_k)
    \]

    \State Construct frozen nonlinear prediction dynamics
    \begin{equation}
\label{eq:frozen_prediction_model}
\dot{\bm x}(t)
=
\bar{\bm f}\!\left(
\bm x(t),\bm u(t);
\bar{\bm N}_v,\bar{\bm N}_w,\bar{\bm B}_v,\bar{\bm B}_w
\right).
\end{equation}

    \State Discretize dynamics using RK4 with step $T_s$ to obtain
    $\bm x_{i+1} = \bm f_d(\bm x_i,\bm u_i)$

    \State Solve the discrete-time OCP
    \[
    \min_{\{\bm u_i\}_{i=0}^{N-1}}
    \sum_{i=0}^{N-1} \ell(\bm x_i,\bm u_i)
    + \ell_f(\bm x_N)
    \]
    subject to dynamics and constraints

    \State Apply first control input $\bm u_k^\star = \bm u_0^\star$

    \State $k \gets k+1$

\EndWhile
\end{algorithmic}
\end{algorithm}

{For practical implementation, the retained modal coordinates
\((\bm\eta,\dot{\bm\eta})\) can be estimated from a finite set of cable measurements rather than obtained by direct distributed-state measurement. A practical sensing configuration may use laser displacement sensors, calibrated camera/vision-based tracking, or compact accelerometers placed at selected cable locations. Laser and camera measurements can provide transverse displacement samples directly, whereas accelerometer signals can be incorporated through filtering or observer-based estimation.}

{Since the reduced cable displacement satisfies \(\bm q(t)=\bm\Phi_r\bm\eta(t)\), the sensor-output model for displacement-type measurements can be written as
\begin{equation}
\bm z_{\mathrm{s}}(t)
=
\bm\Phi_{\mathrm{s}}\bm\eta(t)
+
\bm\varepsilon_{\mathrm{s}}(t),
\label{eq:modal_sensor_model}
\end{equation}
where \(\bm z_{\mathrm{s}}(t)\) is the vector of measured cable displacements, \(\bm\Phi_{\mathrm{s}}\) is the retained modal matrix evaluated at the sensor locations, and \(\bm\varepsilon_{\mathrm{s}}(t)\) denotes measurement noise. Provided that the retained modes are observable from the selected sensor locations, the modal coordinates can be reconstructed through the least-squares projection
\begin{equation}
\hat{\bm\eta}(t)
=
\bm\Phi_{\mathrm{s}}^{\dagger}
\bm z_{\mathrm{s}}(t),
\label{eq:modal_coordinate_estimate}
\end{equation}
where \((\cdot)^{\dagger}\) denotes the Moore--Penrose pseudo-inverse~\cite{GolubVanLoan2013}.}

{The modal velocities \(\dot{\bm\eta}\) can be obtained by filtered differentiation of \(\hat{\bm\eta}\), or more robustly through a reduced-order state estimator based on the modal-reduced cable dynamics, such as a Kalman-type observer or moving-horizon estimator~\cite{Rawlings2017}. Since the present controller retains only \(r=2\) dominant cable modes, the estimator dimension remains small compared with the full finite-element model. Hence, the controller requires estimation of the retained modal states only, rather than reconstruction of the full distributed cable state.}


\section{Numerical Case Studies and Closed-Loop Results}
\label{sec:results_cable_nmpc}

\subsection{Simulation environment and common parameters}
All simulations were performed in \textsc{MATLAB}~R2019b using
\textsc{CasADi}~v3.5.5, an open-source framework for nonlinear optimization and
algorithmic differentiation, with \textsc{Ipopt} employed as the nonlinear
programming solver.
The computations were executed on a standard laptop equipped with an
Intel\textsuperscript{\textregistered} Core\texttrademark\ i7--7660U processor
and 16~GB RAM,
{providing a common numerical prototyping platform for comparing the computational performance of the frozen and non-frozen NMPC formulations.}

All scenarios share the same full nonlinear, modal-reduced vehicle--cable
prediction model and the same NMPC formulation derived in the preceding
sections.
Unless explicitly stated otherwise, the physical parameters, discretization
settings, and NMPC horizon parameters reported in
Table~\ref{tab:common_params} are held fixed across all numerical studies.
Consequently, differences between scenarios arise solely from the enabled plant
disturbance channels, constraint tightening, and/or NMPC weight retuning.

\begin{table}[t]
\centering
\caption{Common system and NMPC parameters used in all scenarios.}
\label{tab:common_params}
\renewcommand{\arraystretch}{1.10}
\setlength{\tabcolsep}{6pt}
\begin{tabular}{ll}
\toprule
Parameter & Value \\
\midrule
Cable span $L$ & $\SI{2.0}{m}$ \\
Cable tension $T$ & $\SI{700}{N}$ \\
Cable elements $n$ & $10$ \\
Retained modes $r$ & $2$ \\
Sampling time  & $\SI{0.02}{s}$ \\
Prediction horizon $N$ & $50$ \\
Simulation time $T_{\mathrm{sim}}$ & $\SI{4.0}{s}$ \\
Initial/goal position  & $(\SI{0.3}{m},\,\SI{1.4}{m})$ \\
Wheel torque limit $\tau_w$ & $\pm\SI{5}{N\,m}$ \\
Arm torque limit $\tau_a$ & $\pm\SI{3}{N\,m}$ \\
\bottomrule
\end{tabular}
\end{table}

\subsection{Scenario description and evaluation protocol}
Six closed-loop scenarios are considered.
Scenario~1 corresponds to nominal disturbance-free traversal.
Scenario~2 considers nominal traversal in the presence of low-amplitude
actuator disturbances, serving as an initial robustness check relative to the disturbance-free case,
Scenario~3 introduces an external cable excitation (point push) to emulate
structural disturbance.
Scenario~4 combines external disturbances with tightened state and input
constraints to assess constraint handling.
Scenario~5 applies stronger disturbances together with retuned NMPC weights to
recover performance.
Scenario~6 corresponds to the most comprehensive disturbance configuration, in
which all disturbance channels are simultaneously enabled.

For each scenario, the reported closed-loop signals include the contact
position $s$ and speed $\dot{s}$, roll and pitch angles $(\phi,\theta)$ and
their rates, representative cable deflection signals reconstructed at the
contact location, and the applied actuator torques $(\tau_w,\tau_a)$.

The NMPC prediction model remains nominal in all cases, while disturbances are
injected exclusively into the plant dynamics and documented through dedicated
disturbance plots (impulse torques, torque ripples, and cable point-push forces).

To ensure clarity and reproducibility, all disturbance channels and mismatch mechanisms
are defined explicitly and applied exclusively to the plant dynamics.
Plant-only actuator disturbances enter additively as
$\tau_w^{\mathrm{plant}}(t)=\tau_w(t)+d_{\tau_w}(t)$ and
$\tau_a^{\mathrm{plant}}(t)=\tau_a(t)+d_{\tau_a}(t)$, where $d_{\tau_w}$ and
$d_{\tau_a}$ consist of a decaying sinusoidal ripple (with frequencies $f_w,f_a$
and decay factor $k_r$) and optional one-sample impulse components.
Roll and pitch disturbances are implemented as equivalent generalized-torque
impulses $\Delta Q_{\phi}=J_{\phi}/T_s$ and $\Delta Q_{\theta}=J_{\theta}/T_s$
applied at prescribed times.
Structural disturbances are modeled through one-sample vertical and lateral
cable point-push impulses of magnitude $J_v/T_s$ and $J_w/T_s$, applied at location
$x_p$ and time $t_0$.
When enabled, plant-model mismatch is introduced by scaling the true plant
parameters relative to the nominal NMPC prediction model as
$\mathcal{T}\mapsto\kappa_{\mathcal{T}}\mathcal{T}$,
$\alpha\mapsto\kappa_{\alpha}\alpha$, and
$\beta\mapsto\kappa_{\beta}\beta$, while the prediction model remains unchanged.

All scenarios are fully deterministic (no randomization); the complete scenario
parameter sets (Tables~\ref{tab:scenarios_dist}--\ref{tab:scenarios_ctrl}),
solver tolerances, and software versions (MATLAB/CasADi/IPOPT) are reported to
enable exact reproduction.

\begin{table}[t]
\centering
\caption{Scenario-wise plant-only disturbance parameters (impulses applied as $J/T_s$).}
\label{tab:scenarios_dist}
\setlength{\tabcolsep}{6pt}
\renewcommand{\arraystretch}{1.15}
\begin{tabular}{lcccccc}
\toprule
Parameter & S1 & S2 & S3 & S4 & S5 & S6 \\
\midrule
$d_{\tau_w}$ amplitude [Nm]     & 0    & 0.50 & 0    & 0.25 & 0.35 & 0.40 \\
$d_{\tau_a}$ amplitude [Nm]     & 0    & -0.35& 0    & -0.20& -0.25& -0.30 \\
Ripple decay $k_r$ [-]          & --   & 6    & --   & 8    & 10   & 10 \\
$f_w$ [Hz]                      & --   & 3.0  & --   & 3.0  & 3.0  & 3.0 \\
$f_a$ [Hz]                      & --   & 2.0  & --   & 2.0  & 2.0  & 2.0 \\
\midrule
Roll impulse time $t_0$ [s]      & --   & 1.00 & --   & 1.00 & 1.00 & 1.00 \\
Roll impulse $J_{\phi}$ [Nms]    & 0    & 0.20 & 0    & 0.12 & 0.15 & 0.15 \\
Pitch impulse time $t_0$ [s]     & --   & 2.00 & --   & 2.00 & 2.00 & 2.00 \\
Pitch impulse $J_{\theta}$ [Nms] & 0    & -0.12& 0    & -0.08& -0.10& -0.10 \\
\midrule
Cable push location $x_p$ [m]    & --   & --   & 1.00 & 1.00 & 1.00 & 1.00 \\
Cable push time $t_0$ [s]        & --   & --   & 1.30 & 1.30 & 1.30 & 1.30 \\
Cable push impulse $J_v$ [Ns]    & 0    & 0    & 0.45 & 0.25 & 0.30 & 0.30 \\
Cable push impulse $J_w$ [Ns]    & 0    & 0    & 0.80 & 0.45 & 0.60 & 0.60 \\
\bottomrule
\end{tabular}
\end{table}

\begin{table}[t]
\centering
\caption{Scenario-wise constraints and NMPC weights (OCP notation). Terminal weights report selected diagonal entries of $\bm P_f$.}
\label{tab:scenarios_ctrl}
\setlength{\tabcolsep}{6pt}
\renewcommand{\arraystretch}{1.15}
\begin{tabular}{lcccccc}
\toprule
Parameter & S1 & S2 & S3 & S4 & S5 & S6 \\
\midrule
$\tau_{w}^{\max}$ [Nm] & 5.0 & 5.0 & 5.0 & 3.0 & 5.0 & 5.0 \\
$\tau_{a}^{\max}$ [Nm] & 3.0 & 3.0 & 3.0 & 1.8 & 3.0 & 3.0 \\
$\phi_{\max}$ [rad]    & 0.5 & 0.5 & 0.5 & 0.25& 0.5 & 0.5 \\
$\theta_{\max}$ [rad]  & 0.5 & 0.5 & 0.5 & 0.35& 0.5 & 0.5 \\
$\Delta \tau_{w}^{\max}$ [Nm/step] & 0.6 & 0.6 & 0.6 & 0.35 & 0.6 & 0.6 \\
$\Delta \tau_{a}^{\max}$ [Nm/step] & 0.4 & 0.4 & 0.4 & 0.25 & 0.4 & 0.4 \\
\midrule
$q_{\eta}$ (with $\bm Q_\eta=q_{\eta}\bm I$) & 80  & 80  & 120 & 80  & 100 & 80 \\
$q_s$      & 140 & 120 & 120 & 120 & 120 & 120 \\
$q_{\dot s}$ & 1.0 & 1.0 & 1.0 & 1.0 & 1.5 & 1.0 \\
$q_{\phi}$   & 30  & 35  & 25  & 45  & 25  & 25 \\
$q_{\theta}$ & 30  & 35  & 25  & 45  & 25  & 25 \\
$q_{\dot\phi}$ & 0.5 & 1.0 & 0.5 & 1.2 & 0.5 & 0.5 \\
$q_{\dot\theta}$& 0.5 & 1.0 & 0.5 & 1.2 & 0.5 & 0.5 \\
$R_w$      & 0.12& 0.18& 0.18& 0.25& 0.22& 0.15 \\
$R_a$      & 0.12& 0.18& 0.18& 0.25& 0.22& 0.15 \\
\midrule
$p_{f,s}$      & 300 & 250 & 250 & 250 & 250 & 250 \\
$p_{f,\phi}$   & 60  & 80  & 60  & 60  & 60  & 60 \\
$p_{f,\theta}$ & 60  & 80  & 60  & 60  & 60  & 60 \\
$p_{f,\eta}$   & 40  & 40  & 70  & 40  & 60  & 40 \\
\midrule
Plant mismatch enabled & no & no & no & no & yes & no \\
$\mathcal{T}$ scale     & -- & -- & -- & -- & 1.20 & -- \\
$\alpha$ scale          & -- & -- & -- & -- & 0.60 & -- \\
$\beta$ scale           & -- & -- & -- & -- & 0.60 & -- \\
\bottomrule
\end{tabular}
\end{table}

\subsection{Reduced-order model accuracy}
\label{subsec:rom_accuracy}

{To assess whether the retained reduced-order model is sufficiently accurate for
control-oriented prediction, a modal-convergence study was performed. The closed-loop
responses obtained with different numbers of retained cable modes were compared against a
higher-order reference model with \(r_{\mathrm{ref}}=6\). The same physical parameters,
scenario settings, controller structure, and disturbance definitions were used, while only
the number of retained cable modes was varied. The comparison was based on the reconstructed
vertical and lateral cable deflections at the moving contact point and at the cable midspan.}

{Table~\ref{tab:rom_accuracy} summarizes the modal-convergence errors for representative
Scenarios~1, 3, and 6. The results show that increasing the number of retained modes reduces
the reconstruction errors relative to the \(r_{\mathrm{ref}}=6\) reference model. The dominant
low-frequency vehicle--cable interaction is captured with a small number of modes, while
higher modes mainly refine localized cable-response components under point-push and
combined-disturbance conditions. The retained \(r=2\) model used in the NMPC therefore provides
a computationally efficient control-oriented approximation, while the modal-convergence study
quantifies the accuracy trade-off associated with modal truncation.}

\begin{table*}[!t]
\centering
\scriptsize
\caption{{Modal-convergence errors of reduced-order cable models relative to the
\(r_{\mathrm{ref}}=6\) reference model. Errors are reported in millimeters.}}
\label{tab:rom_accuracy}
\renewcommand{\arraystretch}{1.12}
\setlength{\tabcolsep}{5.0pt}
\resizebox{\textwidth}{!}{%
\begin{tabular}{clccccccc}
\toprule
Sc. & Case & \(r\) &
RMS \(v_c\) & Max \(v_c\) &
RMS \(w_c\) & Max \(w_c\) &
RMS \(v_m\) & Max \(v_m\) \\
\midrule
1 & Nominal & 1 & 4.413 & 6.872 & 0.677 & 0.993 & 1.566 & 2.461 \\
1 & Nominal & 2 & 4.388 & 6.855 & \(2.75{\times}10^{-15}\) & \(4.58{\times}10^{-15}\) & 1.566 & 2.412 \\
1 & Nominal & 4 & 1.302 & 2.400 & \(6.65{\times}10^{-16}\) & \(1.18{\times}10^{-15}\) & 1.566 & 2.412 \\
1 & Nominal & 5 & 1.302 & 2.400 & 0.0089 & 0.0165 & 1.566 & 2.412 \\
\midrule
3 & Cable push & 1 & 4.448 & 7.341 & 9.196 & 28.10 & 1.672 & 4.370 \\
3 & Cable push & 2 & 4.395 & 6.855 & 0.242 & 2.190 & 1.594 & 3.532 \\
3 & Cable push & 4 & 1.312 & 2.400 & 0.242 & 2.190 & 1.594 & 3.532 \\
3 & Cable push & 5 & 1.311 & 2.400 & 0.0090 & 0.0165 & 1.593 & 3.506 \\
\midrule
6 & All disturbances & 1 & 4.448 & 7.341 & 9.196 & 28.10 & 1.672 & 4.370 \\
6 & All disturbances & 2 & 4.395 & 6.855 & 0.242 & 2.190 & 1.594 & 3.532 \\
6 & All disturbances & 4 & 1.312 & 2.400 & 0.242 & 2.190 & 1.594 & 3.532 \\
6 & All disturbances & 5 & 1.311 & 2.400 & 0.0090 & 0.0165 & 1.593 & 3.506 \\
\bottomrule
\end{tabular}%
}
\end{table*}

{The error magnitudes are small compared with the closed-loop cable deflection levels
reported in the scenario responses, and the convergence trend confirms that the selected
modal truncation captures the dominant cable dynamics relevant to the NMPC controller.}

\subsection{Closed-loop responses across scenarios}

{Figures~\ref{fig:scenarios_grid_13} and~\ref{fig:scenarios_grid_46}
summarize the closed-loop responses of the frozen and non-frozen NMPC
formulations together with the corresponding plant disturbance inputs. For each
scenario, the left panel compares the frozen and non-frozen closed-loop
trajectories, including the main states and applied inputs, while the right
panel reports the disturbance signals injected into the nonlinear plant.}

{The disturbance plots are included to make the excitation content explicit and
to confirm that the NMPC prediction model remains nominal while disturbances are
applied only to the plant dynamics.}

\subsection{Quantitative performance metrics}

{The non-frozen NMPC formulation is used as a direct comparative baseline for
evaluating the proposed frozen-interpolation strategy. In this baseline, the
same reduced-order nonlinear vehicle--cable model, objective function,
constraints, physical parameters, and scenario definitions are retained, while
the contact-point interpolation and slope operators are re-evaluated along the
prediction horizon according to the predicted contact positions. Thus, the
comparison isolates the effect of freezing the spatial contact operators while
keeping the remaining NMPC formulation unchanged.}

\begin{figure*}[p]
\centering
\captionsetup{font=small}
\setlength{\abovecaptionskip}{4pt}
\setlength{\belowcaptionskip}{0pt}

\begin{subfigure}[t]{0.50\textwidth}
  \centering
  \includegraphics[width=\linewidth]{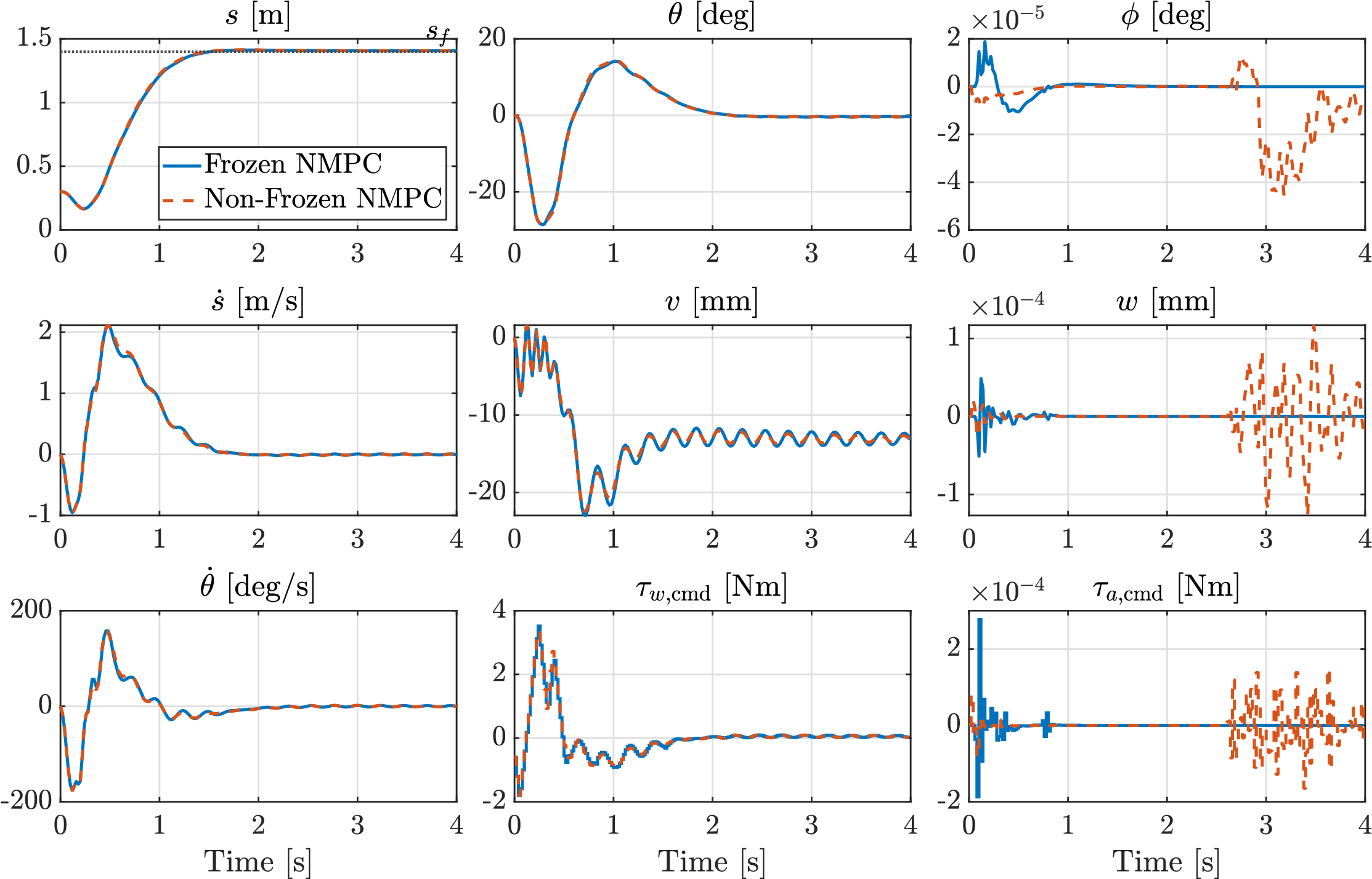}
  \caption{Scenario~1: frozen versus non-frozen closed-loop response.}
  \label{fig:scen1_resp}
\end{subfigure}\hfill
\begin{subfigure}[t]{0.50\textwidth}
  \centering
  \includegraphics[width=\linewidth]{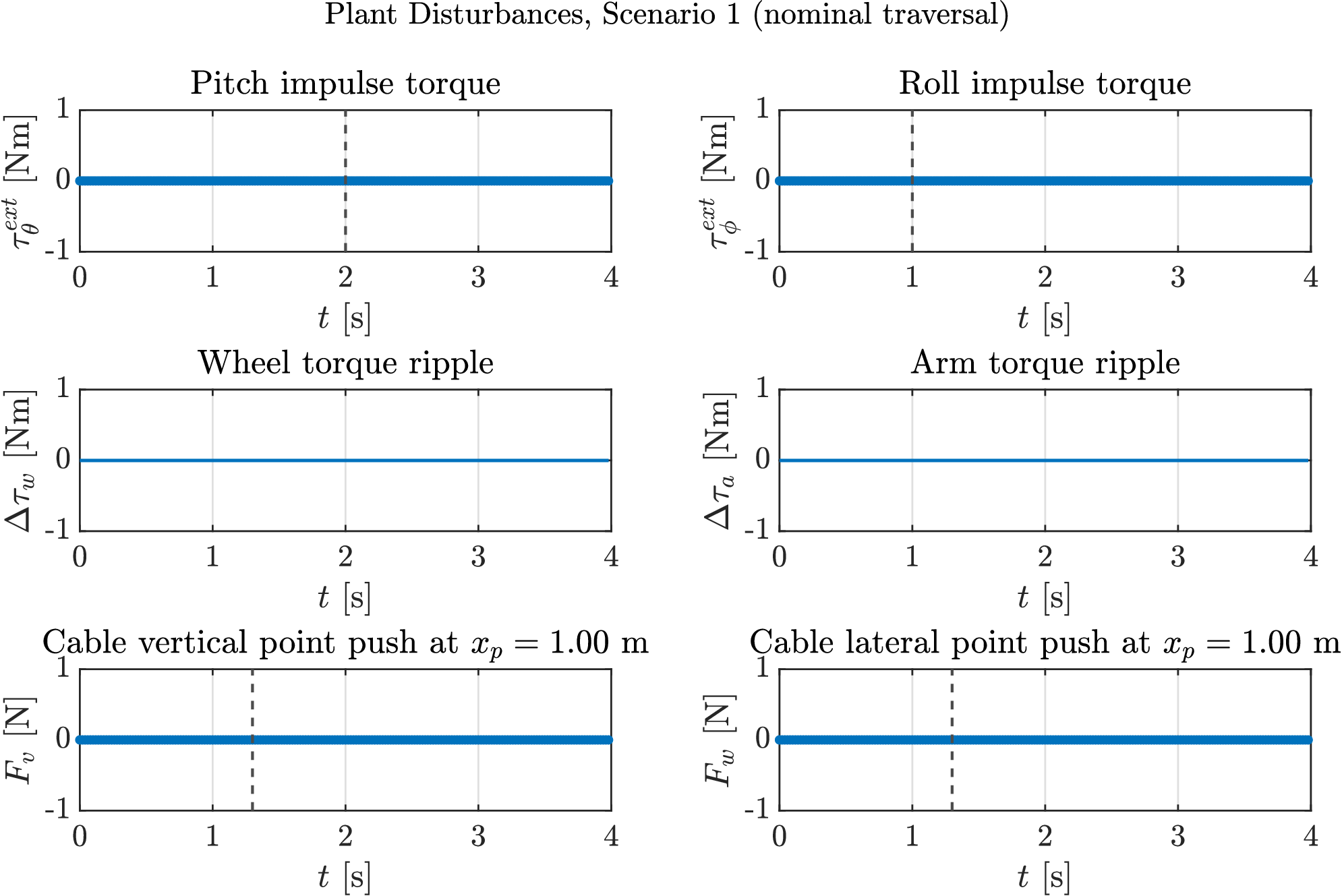}
  \caption{Scenario~1: plant disturbance inputs.}
  \label{fig:scen1_dist}
\end{subfigure}

\vspace{2.5mm}

\begin{subfigure}[t]{0.50\textwidth}
  \centering
  \includegraphics[width=\linewidth]{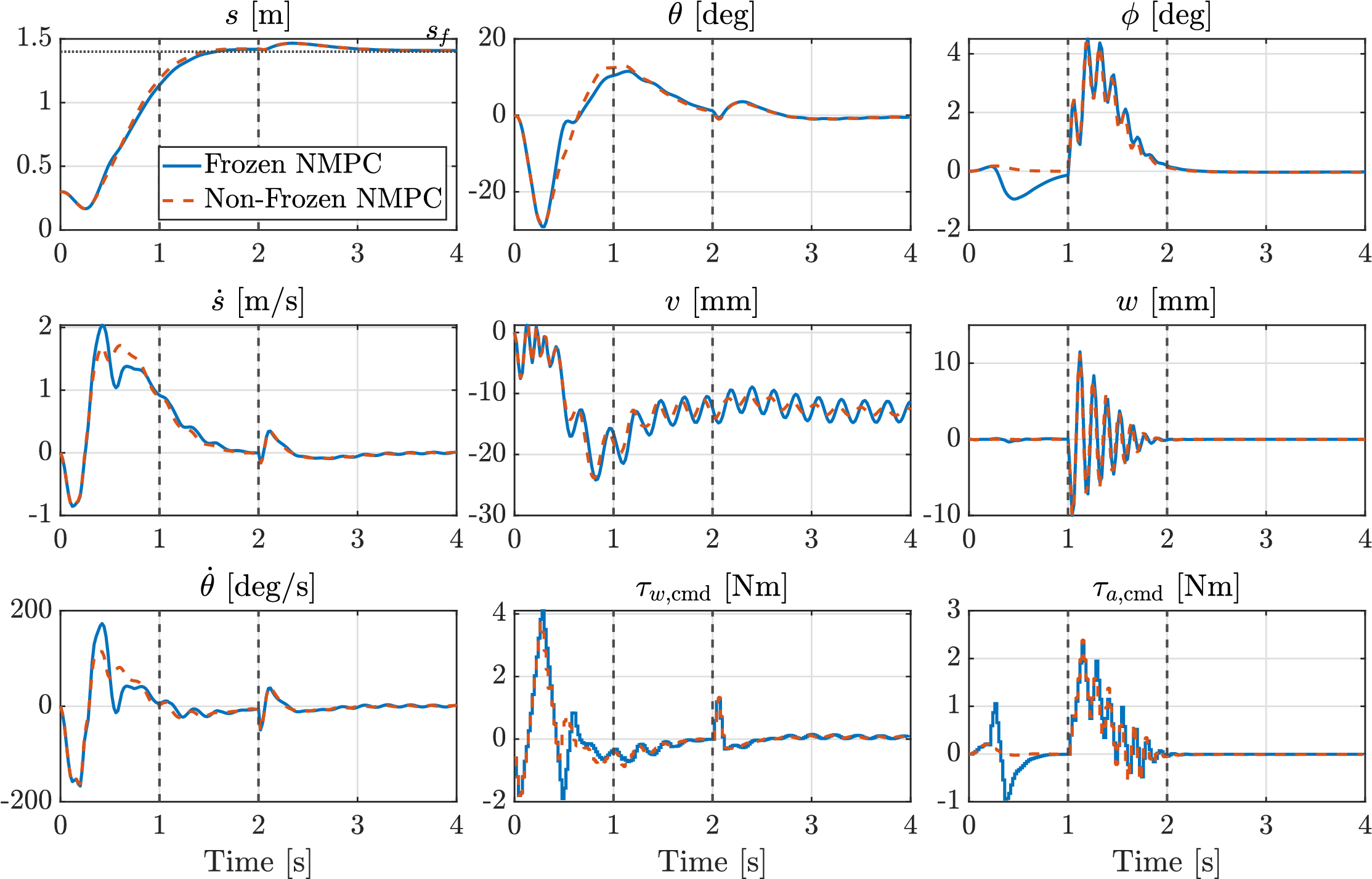}
  \caption{Scenario~2: frozen versus non-frozen closed-loop response.}
  \label{fig:scen2_resp}
\end{subfigure}\hfill
\begin{subfigure}[t]{0.50\textwidth}
  \centering
  \includegraphics[width=\linewidth]{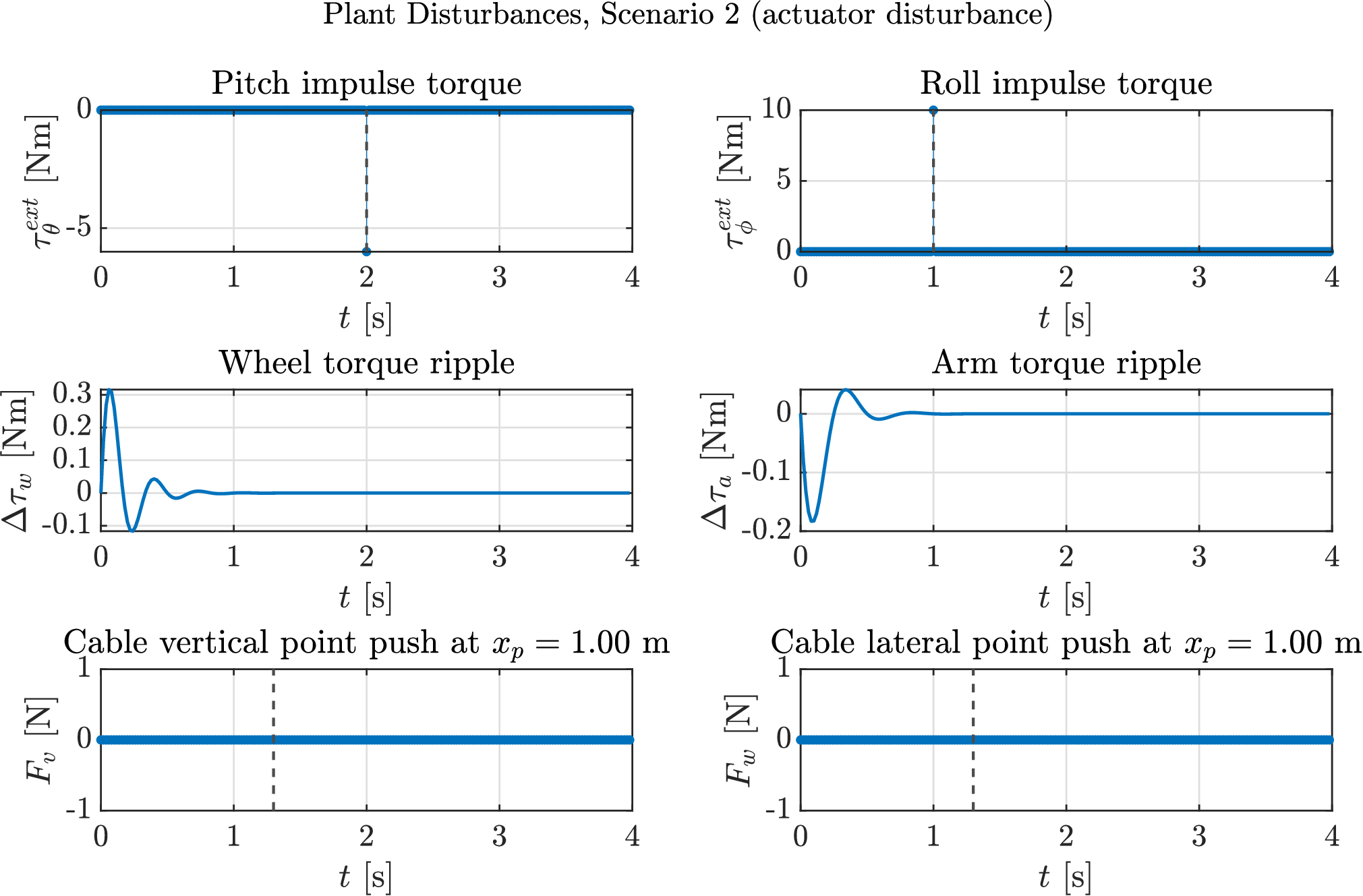}
  \caption{Scenario~2: plant disturbance inputs.}
  \label{fig:scen2_dist}
\end{subfigure}

\vspace{2.5mm}

\begin{subfigure}[t]{0.50\textwidth}
  \centering
  \includegraphics[width=\linewidth]{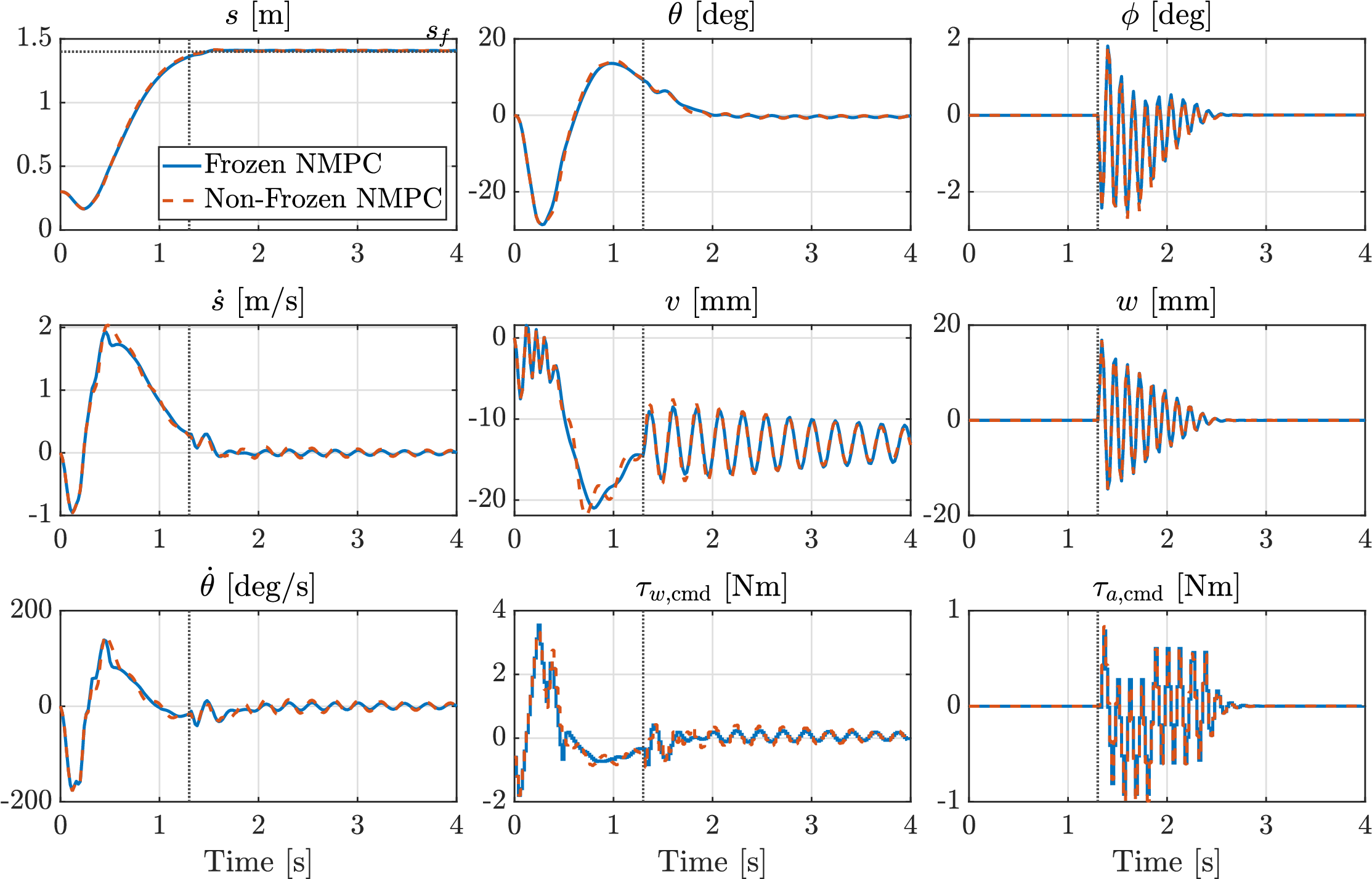}
  \caption{Scenario~3: frozen versus non-frozen closed-loop response.}
  \label{fig:scen3_resp}
\end{subfigure}\hfill
\begin{subfigure}[t]{0.50\textwidth}
  \centering
  \includegraphics[width=\linewidth]{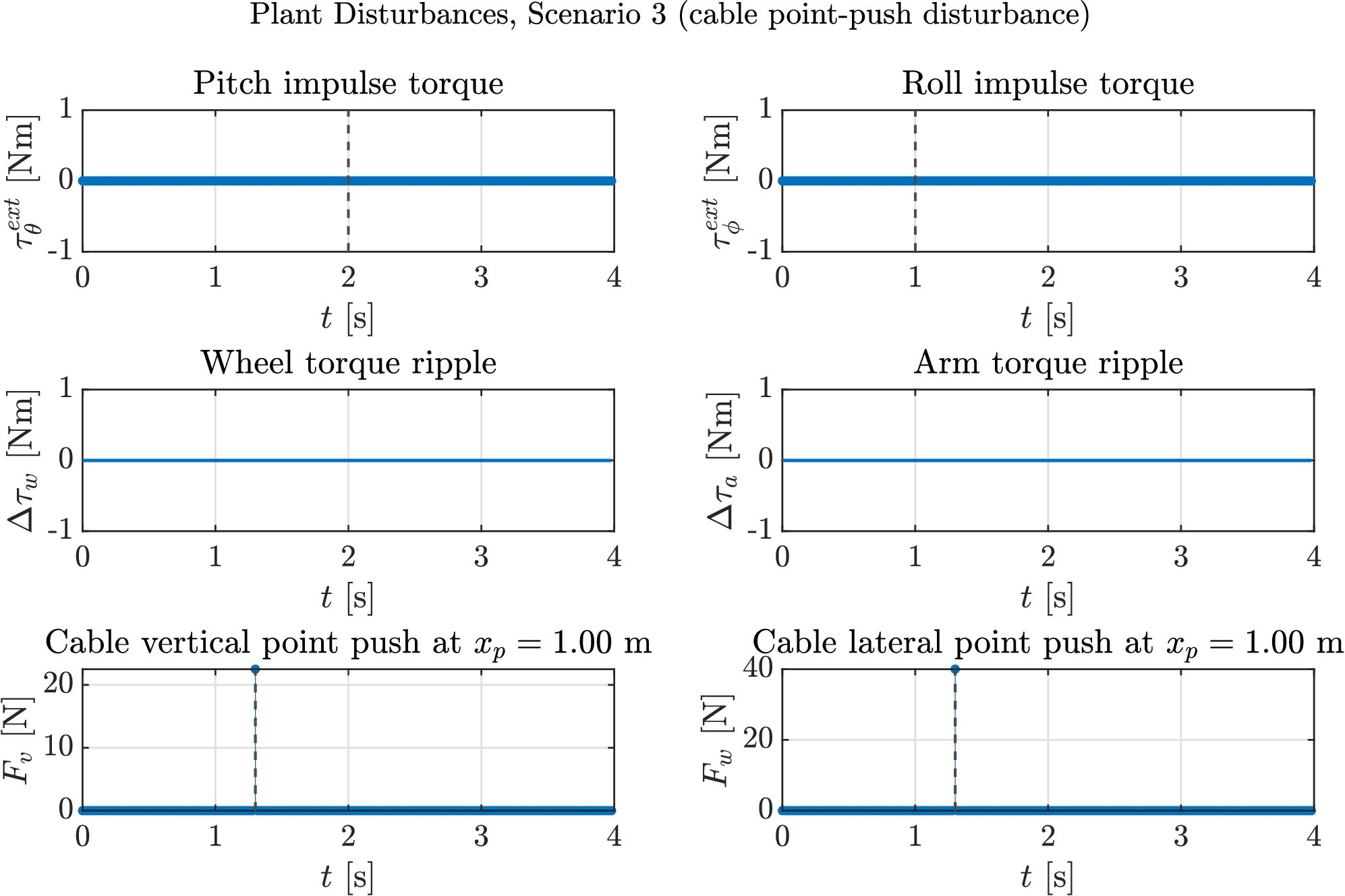}
  \caption{Scenario~3: plant disturbance inputs.}
  \label{fig:scen3_dist}
\end{subfigure}

\caption{{Closed-loop frozen-versus-non-frozen NMPC responses (left) and corresponding plant disturbance inputs (right) for Scenarios~1--3.}}
\label{fig:scenarios_grid_13}
\end{figure*}

\begin{figure*}[p]
\centering
\captionsetup{font=small}
\setlength{\abovecaptionskip}{4pt}
\setlength{\belowcaptionskip}{0pt}

\begin{subfigure}[t]{0.50\textwidth}
  \centering
  \includegraphics[width=\linewidth]{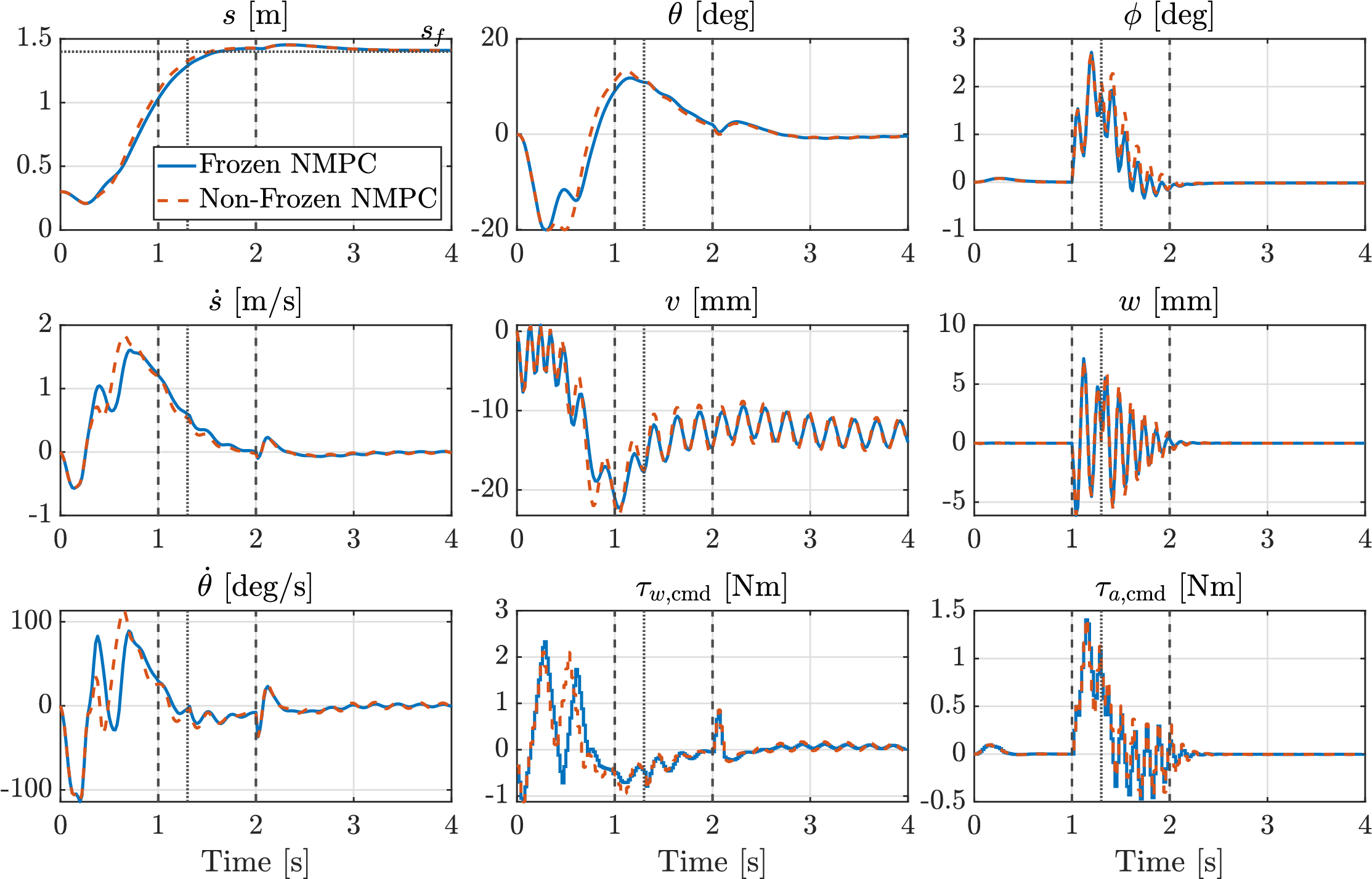}
  \caption{Scenario~4: frozen versus non-frozen closed-loop response.}
  \label{fig:scen4_resp}
\end{subfigure}\hfill
\begin{subfigure}[t]{0.50\textwidth}
  \centering
  \includegraphics[width=\linewidth]{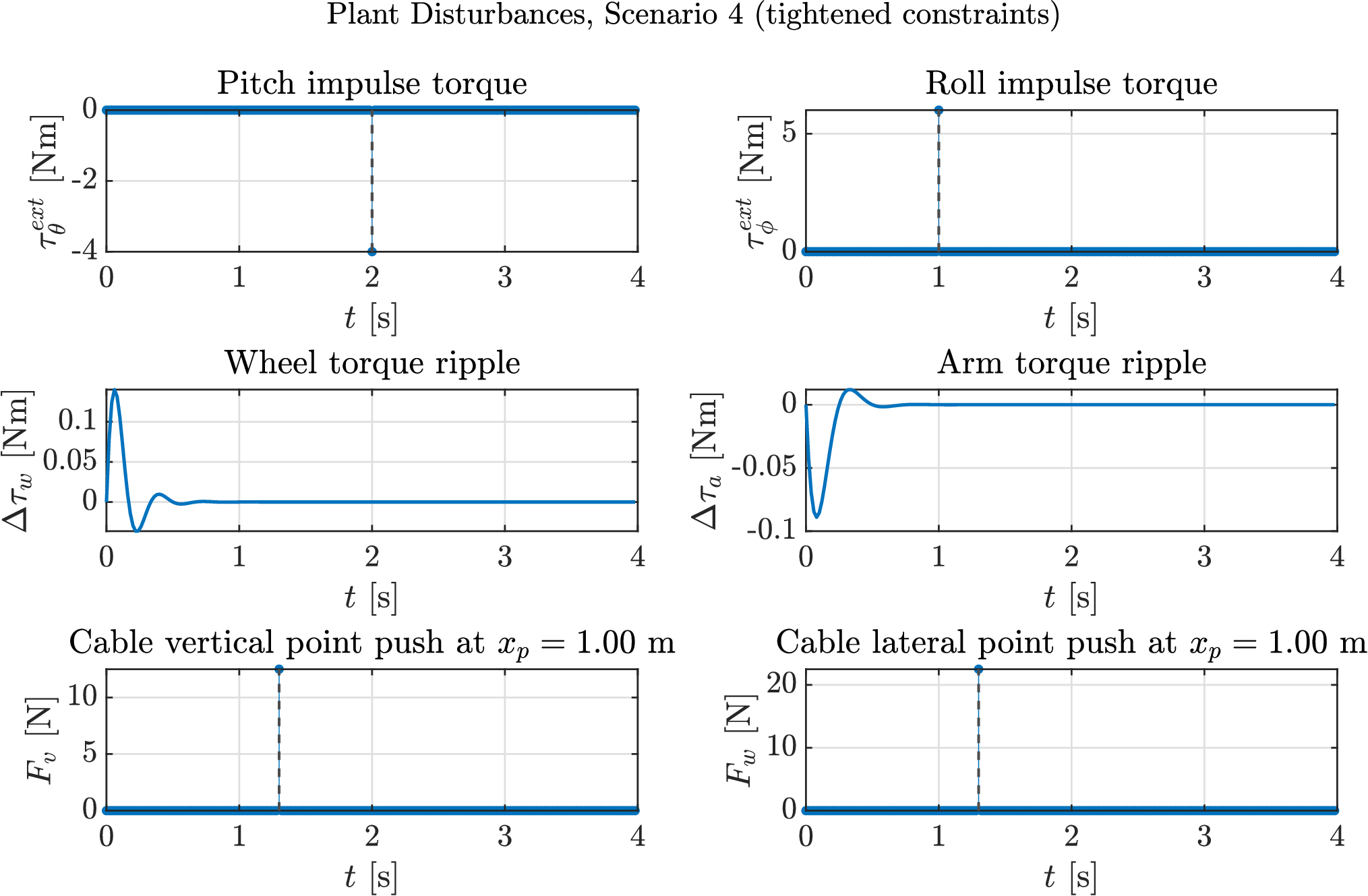}
  \caption{Scenario~4: plant disturbance inputs.}
  \label{fig:scen4_dist}
\end{subfigure}

\vspace{2.5mm}

\begin{subfigure}[t]{0.50\textwidth}
  \centering
  \includegraphics[width=\linewidth]{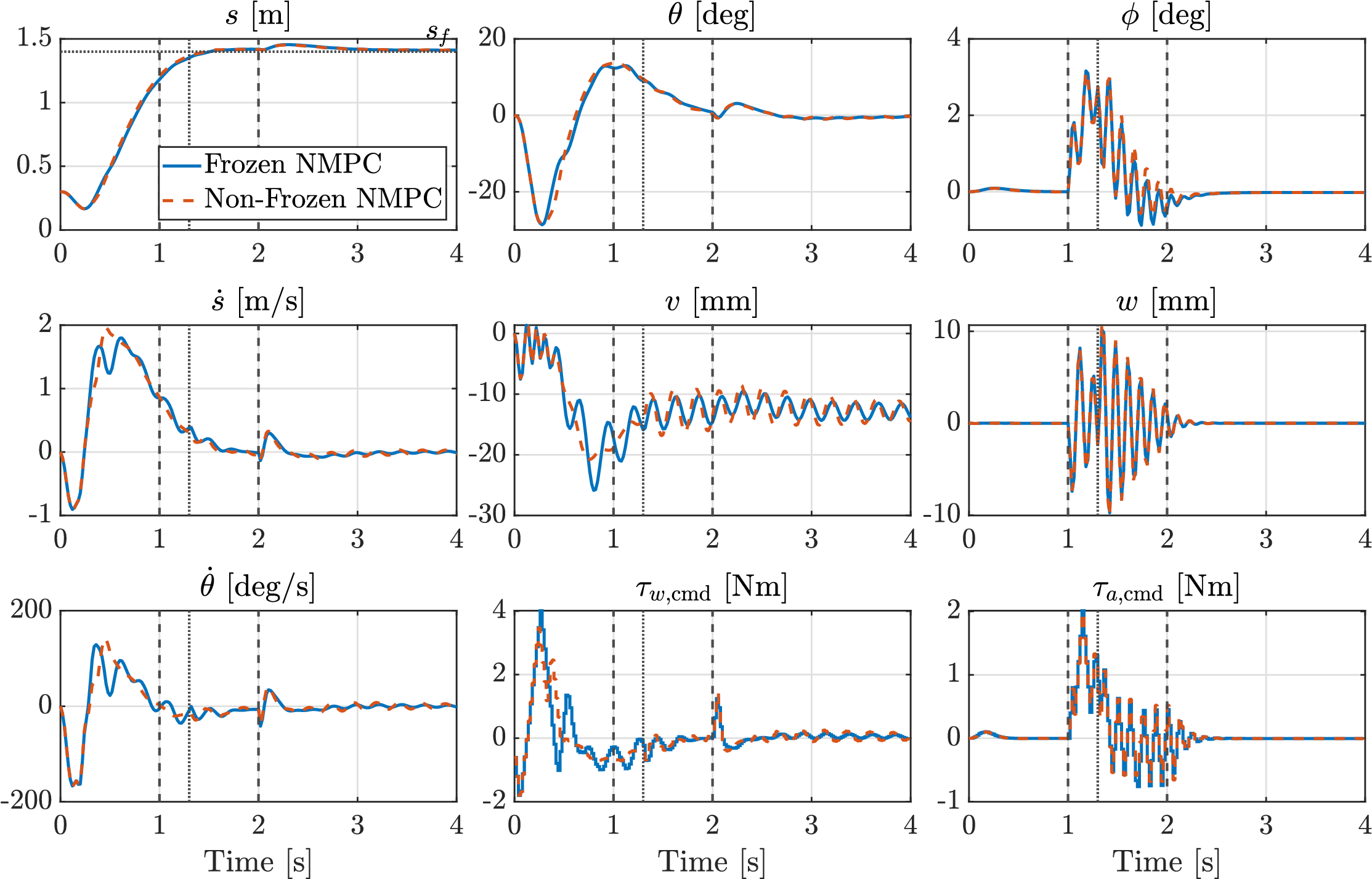}
  \caption{Scenario~5: frozen versus non-frozen closed-loop response.}
  \label{fig:scen5_resp}
\end{subfigure}\hfill
\begin{subfigure}[t]{0.50\textwidth}
  \centering
  \includegraphics[width=\linewidth]{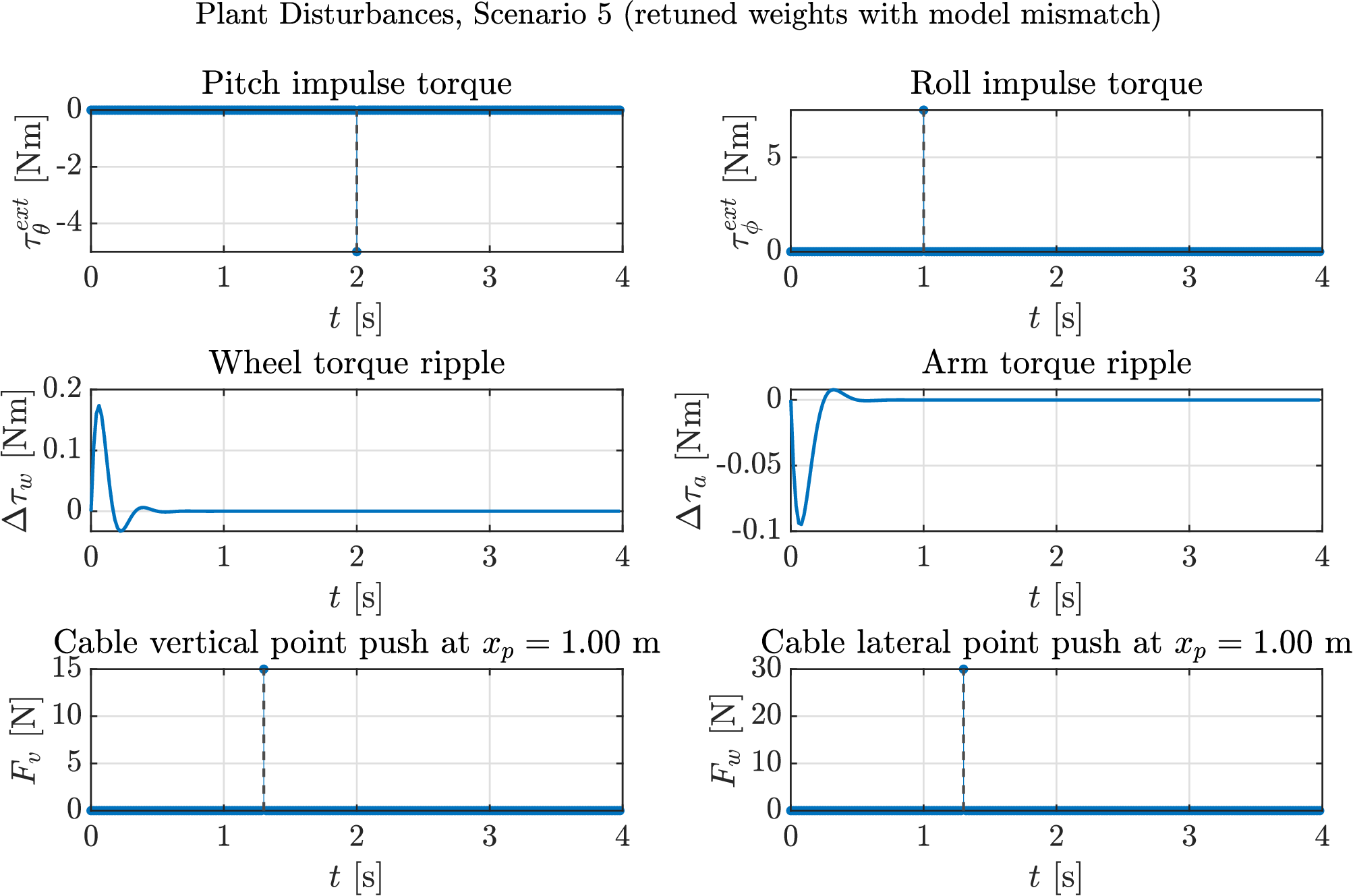}
  \caption{Scenario~5: plant disturbance inputs.}
  \label{fig:scen5_dist}
\end{subfigure}

\vspace{2.5mm}

\begin{subfigure}[t]{0.50\textwidth}
  \centering
  \includegraphics[width=\linewidth]{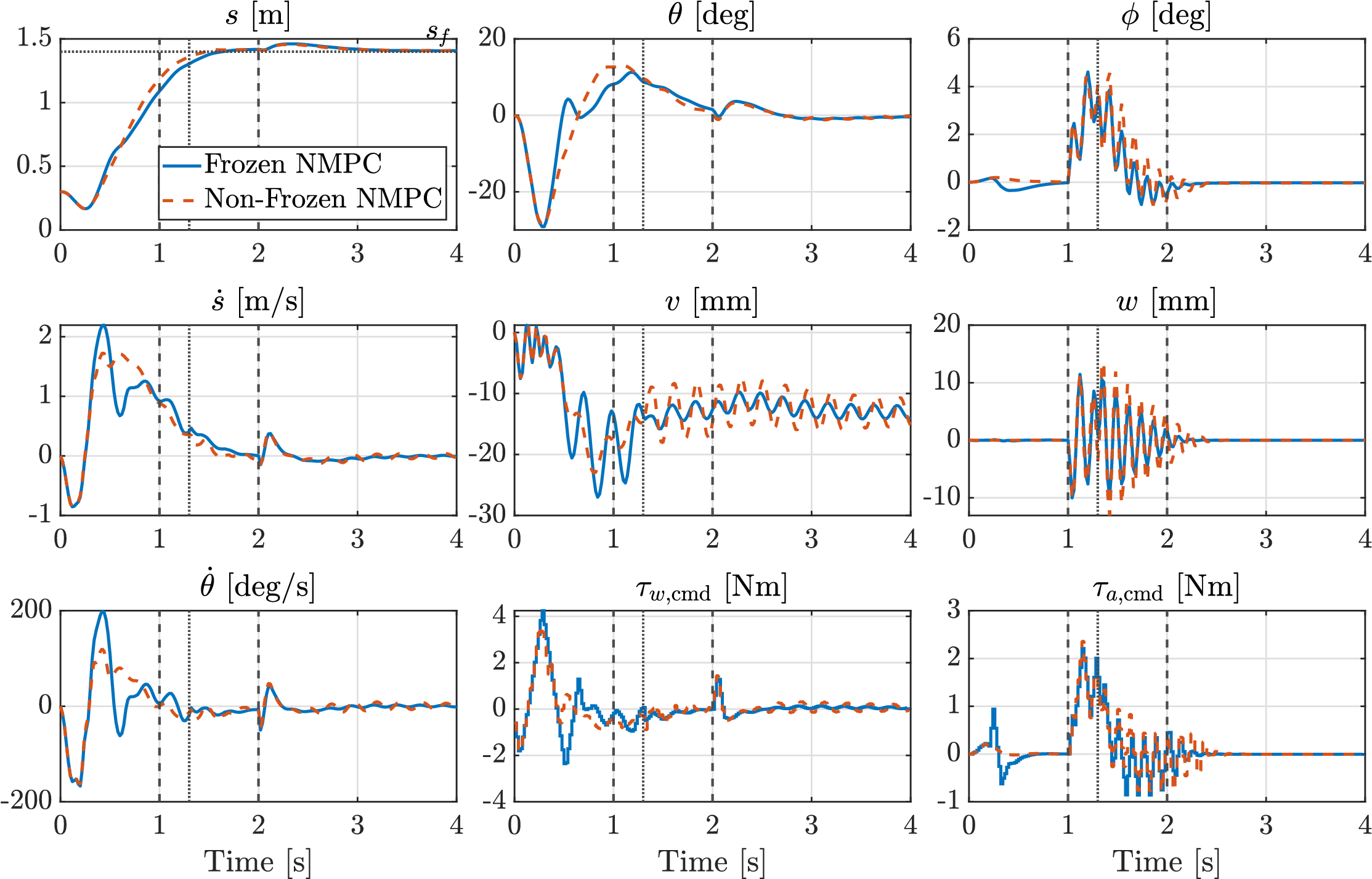}
  \caption{Scenario~6: frozen versus non-frozen closed-loop response.}
  \label{fig:scen6_resp}
\end{subfigure}\hfill
\begin{subfigure}[t]{0.50\textwidth}
  \centering
  \includegraphics[width=\linewidth]{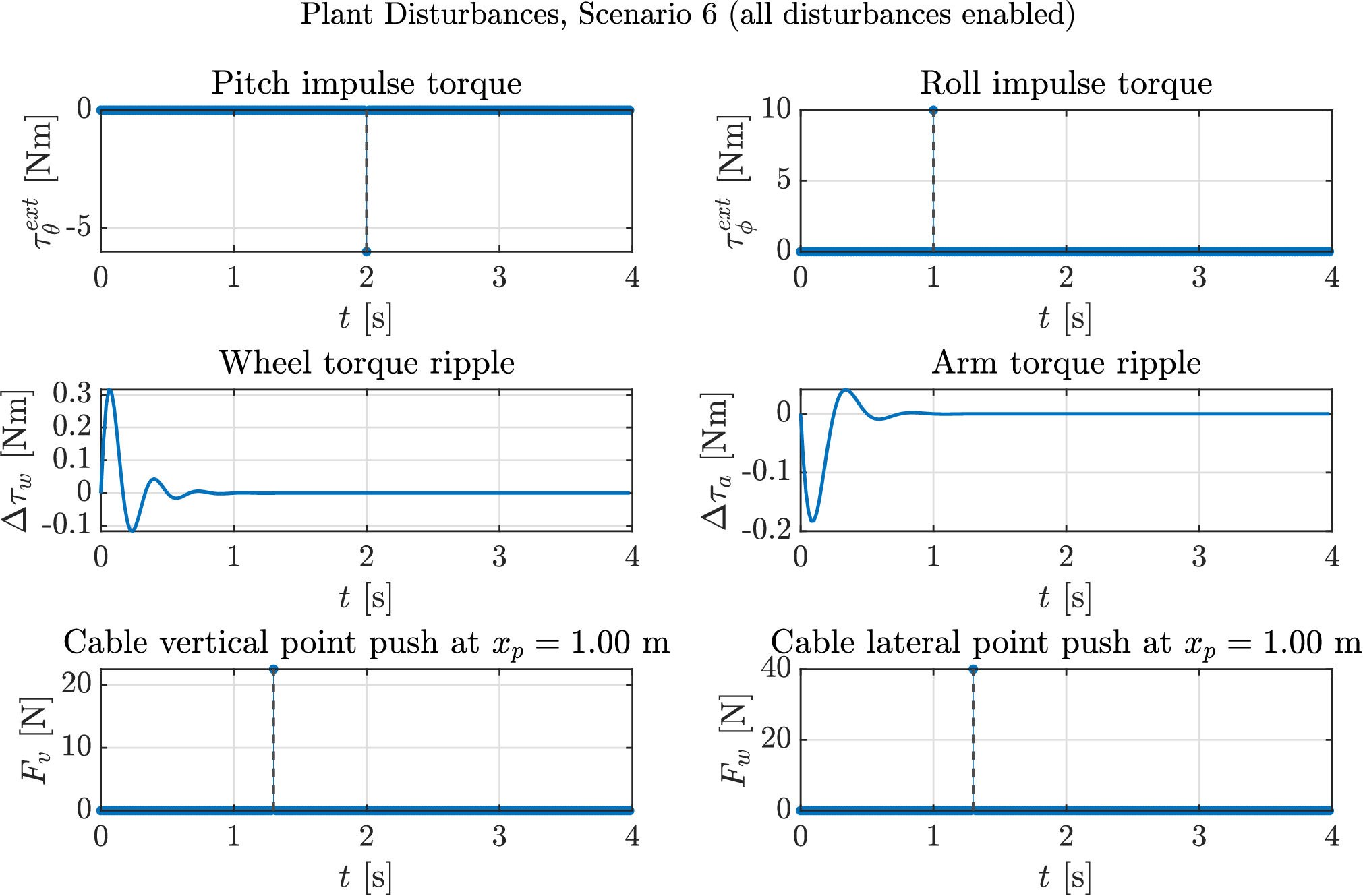}
  \caption{Scenario~6: plant disturbance inputs.}
  \label{fig:scen6_dist}
\end{subfigure}

\caption{{Closed-loop frozen-versus-non-frozen NMPC responses (left) and corresponding plant disturbance inputs (right) for Scenarios~4--6.}}
\label{fig:scenarios_grid_46}
\end{figure*}

{Table~\ref{tab:scenario_metrics} shows that the frozen and non-frozen
formulations provide comparable closed-loop tracking, attitude-regulation,
cable-response, and actuator-use behavior across the tested scenarios.
Differences between the two formulations become more visible under actuator
disturbances, tightened constraints, model mismatch, and combined disturbance
channels, but both methods maintain the same qualitative closed-loop behavior.
The quantitative metrics in Table~\ref{tab:scenario_metrics} are consistent
with the trajectory comparisons in Figs.~\ref{fig:scenarios_grid_13}
and~\ref{fig:scenarios_grid_46}.}

{These results indicate that the frozen-interpolation formulation preserves the
main closed-loop performance of the non-frozen NMPC baseline. The corresponding
computational effect of freezing the contact operators is quantified later in
the timing analysis, where the solver times of the frozen and non-frozen
formulations are compared directly.}
\begin{table*}[!t]
\centering
\scriptsize
\caption{{Closed-loop performance comparison between the frozen-interpolation
NMPC formulation and the non-frozen NMPC baseline across Scenarios~1--6.
The final position error is \(e_s=s(T_{\mathrm{sim}})-s_f\), angles are reported
in degrees, and cable deflections are reported at the moving contact point.}}
\label{tab:scenario_metrics}
\renewcommand{\arraystretch}{1.12}

\textbf{(a) Tracking and attitude-performance metrics}

\vspace{1mm}
\setlength{\tabcolsep}{6.0pt}
\begin{tabular}{cllcccc}
\toprule
Sc. & Scenario & Method &
\(e_s\) [m] &
RMS \(e\) [m] &
\(|\phi|_{\max}\) / RMS [deg] &
\(|\theta|_{\max}\) / RMS [deg] \\
\midrule
1 & Nominal & Frozen & \(+0.00597\) & 0.4493 & 0.00003 / 0.00001 & 28.63 / 8.77 \\
1 & Nominal & NonFrozen & \(+0.00619\) & 0.4487 & 0.00005 / 0.00001 & 28.65 / 8.82 \\
\midrule
2 & Actuator disturbance & Frozen & \(+0.00906\) & 0.4618 & 4.665 / 1.133 & 28.96 / 7.35 \\
2 & Actuator disturbance & NonFrozen & \(+0.00865\) & 0.4539 & 4.667 / 1.143 & 28.66 / 8.37 \\
\midrule
3 & Cable point push & Frozen & \(+0.00649\) & 0.4490 & 3.110 / 0.764 & 28.63 / 8.87 \\
3 & Cable point push & NonFrozen & \(+0.00685\) & 0.4487 & 2.541 / 0.593 & 28.65 / 8.83 \\
\midrule
4 & Tightened constraints & Frozen & \(+0.00991\) & 0.4827 & 2.850 / 0.609 & 20.06 / 7.54 \\
4 & Tightened constraints & NonFrozen & \(+0.00993\) & 0.4804 & 2.839 / 0.588 & 20.05 / 7.90 \\
\midrule
5 & Retuned mismatch & Frozen & \(+0.01224\) & 0.4520 & 3.224 / 0.701 & 28.65 / 8.61 \\
5 & Retuned mismatch & NonFrozen & \(+0.01264\) & 0.4510 & 3.208 / 0.717 & 28.65 / 8.79 \\
\midrule
6 & All disturbances & Frozen & \(+0.00681\) & 0.4625 & 4.675 / 1.025 & 29.00 / 7.41 \\
6 & All disturbances & NonFrozen & \(+0.00796\) & 0.4535 & 4.685 / 1.078 & 28.66 / 8.48 \\
\bottomrule
\end{tabular}

\vspace{3mm}

\textbf{(b) Cable-response and actuator-use metrics}

\vspace{1mm}
\setlength{\tabcolsep}{7.0pt}
\begin{tabular}{cllccc}
\toprule
Sc. & Scenario & Method &
\(|v|_{\max}\) / RMS [m] &
\(|w|_{\max}\) / RMS [m] &
RMS \(\tau_w\) / \(\tau_a\) [Nm] \\
\midrule
1 & Nominal & Frozen
& 0.0220 / 0.0142
& \(1.31{\times}10^{-7}\) / \(2.21{\times}10^{-8}\)
& 0.696 / 0.000 \\
1 & Nominal & NonFrozen
& 0.0227 / 0.0142
& \(1.43{\times}10^{-7}\) / \(2.42{\times}10^{-8}\)
& 0.693 / 0.000 \\
\midrule
2 & Actuator disturbance & Frozen
& 0.0361 / 0.0159 & 0.0117 / 0.00220 & 0.873 / 0.465 \\
2 & Actuator disturbance & NonFrozen
& 0.0262 / 0.0139 & 0.0117 / 0.00213 & 0.684 / 0.454 \\
\midrule
3 & Cable point push & Frozen
& 0.0225 / 0.0143 & 0.0174 / 0.00374 & 0.717 / 0.271 \\
3 & Cable point push & NonFrozen
& 0.0226 / 0.0142 & 0.0174 / 0.00373 & 0.709 / 0.236 \\
\midrule
4 & Tightened constraints & Frozen
& 0.0271 / 0.0138 & 0.00699 / 0.00158 & 0.522 / 0.281 \\
4 & Tightened constraints & NonFrozen
& 0.0257 / 0.0137 & 0.00689 / 0.00156 & 0.512 / 0.283 \\
\midrule
5 & Retuned mismatch & Frozen
& 0.0237 / 0.0139 & 0.00946 / 0.00222 & 0.682 / 0.370 \\
5 & Retuned mismatch & NonFrozen
& 0.0228 / 0.0138 & 0.00989 / 0.00231 & 0.696 / 0.373 \\
\midrule
6 & All disturbances & Frozen
& 0.0369 / 0.0161 & 0.0123 / 0.00290 & 0.914 / 0.505 \\
6 & All disturbances & NonFrozen
& 0.0248 / 0.0139 & 0.0116 / 0.00299 & 0.688 / 0.485 \\
\bottomrule
\end{tabular}
\end{table*}

\subsection{Frozen-versus-non-frozen prediction mismatch and speed sensitivity}
\label{subsec:frozen_nonfrozen_prediction_mismatch}

{To verify that the computational benefit of freezing the spatial contact-point interpolation operators is obtained without introducing an unquantified prediction error, a one-horizon prediction-mismatch study was performed between the proposed frozen NMPC prediction model and a non-frozen prediction model with horizon-wise contact-operator updates.}

{At selected closed-loop horizon-starting times \(t_h\), both prediction models were initialized from the same closed-loop state and propagated over one NMPC horizon using the same input sequence. In the frozen model, the reduced interpolation and slope operators \(N_v\), \(N_w\), \(B_v\), and \(B_w\) were evaluated once at \(s(t_h)\) and held fixed over the horizon. In the non-frozen model, the same operators were re-evaluated at the predicted contact position at every prediction step. Thus, the reported mismatch isolates the effect of the frozen-versus-non-frozen treatment of the spatial contact operators.}

{For each state or reconstructed cable-response quantity \(z\), the mean absolute one-horizon mismatch was computed as}
{
\begin{equation}
\overline{e}_z(t_h)
=
\frac{1}{N+1}
\sum_{i=0}^{N}
\left|
z_{\mathrm{fr}}(i;t_h)-z_{\mathrm{nf}}(i;t_h)
\right|,
\label{eq:mean_one_horizon_mismatch}
\end{equation}
}
{where \(z_{\mathrm{fr}}\) and \(z_{\mathrm{nf}}\) denote the frozen and non-frozen predictions, respectively, \(i\) is the prediction-step index, and \(N\) is the NMPC prediction horizon.}

\begin{table}[!t]
\centering
\scriptsize
\caption{{Mean one-horizon prediction mismatch between frozen and non-frozen contact-point interpolation models. The notation \(\overline{e}_z\) denotes the mean absolute prediction mismatch of variable \(z\) over one NMPC horizon. Both models are initialized from the same closed-loop state at horizon-start time \(t_h\) and propagated using the same input sequence.}}
\label{tab:table_predictionmismatch_frozen_vs_nonfrozen}
\resizebox{\textwidth}{!}{%
\begin{tabular}{cccccccccc}
\hline
Scenario &
\(t_h\) [s] &
\(\overline{e}_s\) [m] &
\(\overline{e}_{\dot{s}}\) [m/s] &
\(\overline{e}_{\phi}\) [deg] &
\(\overline{e}_{\theta}\) [deg] &
\(\overline{e}_{\eta}\) &
\(\overline{e}_{\dot{\eta}}\) &
\(\overline{e}_{v}\) [m] &
\(\overline{e}_{w}\) [m] \\
\hline
1 & 0.2 & 0.01672 & 0.09588 & \(3.444\times10^{-4}\) & 4.956 & 0.002945 & 0.01753 & 0.008287 & \(2.535\times10^{-7}\) \\
1 & 1.0 & 0.02857 & 0.1607 & \(5.656\times10^{-6}\) & 8.662 & 0.001436 & 0.005288 & 0.003907 & \(2.482\times10^{-8}\) \\
1 & 2.0 & \(7.656\times10^{-4}\) & 0.005171 & \(1.301\times10^{-9}\) & 0.1937 & \(3.569\times10^{-5}\) & \(2.783\times10^{-4}\) & \(8.381\times10^{-5}\) & \(2.228\times10^{-13}\) \\
3 & 0.2 & 0.01732 & 0.09866 & \(1.618\times10^{-10}\) & 5.194 & 0.002897 & 0.02013 & 0.008144 & \(9.329\times10^{-15}\) \\
3 & 1.0 & 0.02813 & 0.1588 & \(6.081\times10^{-13}\) & 8.549 & 0.001428 & 0.005645 & 0.003868 & \(1.552\times10^{-16}\) \\
3 & 2.0 & \(3.571\times10^{-4}\) & 0.002649 & 0.1821 & 0.08574 & \(9.045\times10^{-5}\) & 0.001632 & \(2.266\times10^{-4}\) & \(3.560\times10^{-5}\) \\
6 & 0.2 & 0.004537 & 0.04775 & 0.04978 & 1.262 & 0.003201 & 0.01747 & 0.009112 & \(9.701\times10^{-6}\) \\
6 & 1.0 & 0.04741 & 0.2162 & 0.01735 & 13.35 & 0.002733 & 0.06820 & 0.007308 & \(3.142\times10^{-6}\) \\
6 & 2.0 & 0.03243 & 0.1733 & 0.002570 & 8.928 & 0.001004 & 0.02430 & 0.002693 & \(5.809\times10^{-7}\) \\
\hline
\end{tabular}%
}
\end{table}

{The resulting mean one-horizon mismatches are summarized in Table~\ref{tab:table_predictionmismatch_frozen_vs_nonfrozen}. The axial-position, roll-angle, and lateral-deflection mismatches remain limited over the tested horizon-starting times, whereas the pitch channel is the most sensitive during transient portions of the maneuver. This behavior is consistent with the inherently unstable pitch dynamics of the electric unicycle and with the open-loop nature of the one-horizon comparison.}

\begin{figure}[!t]
\centering
\includegraphics[width=0.95\linewidth]{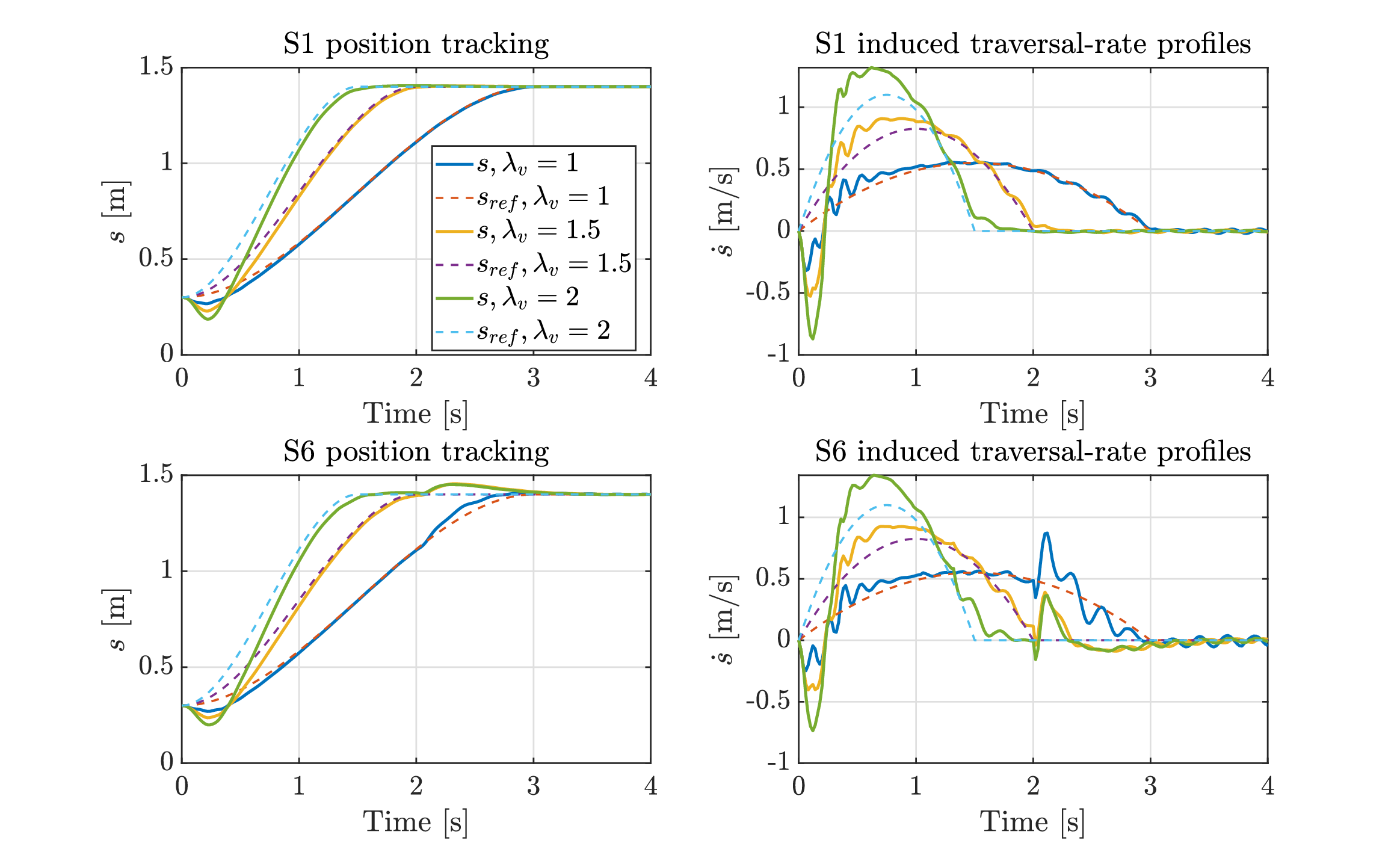}
\caption{{Speed-scaled reference and realized traversal-rate profiles used to assess the frozen-interpolation approximation under faster contact-point migration. Scenario~1 represents nominal traversal, while Scenario~6 represents the most demanding all-disturbance case. The dimensionless traversal-speed scale \(\lambda_v\) modifies the reference motion, while the physical model, sampling time, actuator limits, and NMPC formulation are kept unchanged.}}
\label{fig:speed_scaling_tracking_profiles} \end{figure}

{To assess how the frozen-interpolation approximation changes under faster contact-point migration, an additional speed-scaling study was performed. The traversal reference was scaled using the dimensionless traversal-speed factor \(\lambda_v=1.0\), \(1.5\), and \(2.0\), where \(\lambda_v=1.0\) denotes the nominal traversal profile. The physical parameters, sampling time, actuator limits, and NMPC formulation were kept unchanged. The study was performed for Scenario~1 and Scenario~6, representing the nominal disturbance-free case and the most demanding all-disturbance case, respectively. Figure~\ref{fig:speed_scaling_tracking_profiles} shows the corresponding speed-scaled reference and realized traversal-rate profiles. For each speed case, the same mean one-horizon mismatch metric defined above was evaluated.}

\begin{table}[!t]
\centering
\scriptsize
\caption{{Effect of traversal-speed scaling on the mean frozen-versus-non-frozen prediction mismatch. The reported values are mean absolute one-horizon errors computed using Eq.~\eqref{eq:mean_one_horizon_mismatch}.}}
\label{tab:table_speedscaling_frozenerror}
\resizebox{\textwidth}{!}{%
\begin{tabular}{ccccccc}
\hline
Speed scale \(\lambda_v\) &
\(\overline{e}_s\) [m] &
\(\overline{e}_{\dot{s}}\) [m/s] &
\(\overline{e}_{\theta}\) [deg] &
\(\overline{e}_{\phi}\) [deg] &
\(\overline{e}_{v}\) [m] &
\(\overline{e}_{w}\) [m] \\
\hline
1.0 & 0.01572 & 0.2561 & 20.09 & \(4.455\times10^{-13}\) & 0.03072 & \(3.265\times10^{-16}\) \\
1.5 & 0.04388 & 0.4606 & 28.20 & \(4.743\times10^{-13}\) & 0.02969 & \(3.867\times10^{-16}\) \\
2.0 & 0.09298 & 0.8838 & 38.93 & \(1.693\times10^{-13}\) & 0.02736 & \(1.581\times10^{-16}\) \\
\hline
\end{tabular}%
}
\end{table}

{The speed-scaling results in Table~\ref{tab:table_speedscaling_frozenerror}, together with Fig.~\ref{fig:speed_scaling_tracking_profiles}, show that the mismatch increases with traversal speed, as expected, because the contact point travels a larger distance over one prediction horizon. In particular, the mean axial-position mismatch increases from \(0.01572\) m at \(\lambda_v=1.0\) to \(0.09298\) m at \(\lambda_v=2.0\), while the mean pitch mismatch increases from \(20.09^\circ\) to \(38.93^\circ\). These results confirm that the frozen-interpolation approximation is speed-dependent. For faster traversal, tighter prediction accuracy can be obtained by reducing the prediction horizon, decreasing the sampling time, or updating the contact operators more frequently. In the receding-horizon implementation used here, the frozen contact operators are re-evaluated at every sampling instant using the updated measured contact position, preventing the one-horizon mismatch from accumulating over the closed-loop simulation.}

\subsection{Computational performance}
\label{subsec:computational_performance}

{To evaluate the computational effect of the frozen spatial contact-point interpolation
strategy, solver timing was recorded for both the frozen and non-frozen NMPC formulations.
The non-frozen formulation recomputes the contact-point interpolation operators at every
prediction step, whereas the frozen formulation evaluates them once at the measured contact
position at the beginning of each sampling instant and holds them fixed over the prediction
horizon.}

{Table~\ref{tab:table_timing_frozen_vs_nonfrozen} reports the mean, median,
95th-percentile, and maximum NMPC solve times for all scenarios. The timing values
correspond to the MATLAB/CasADi/IPOPT implementation executed on the reported Intel Core
i7 laptop. These timings therefore characterize a numerical prototyping benchmark rather
than certified embedded real-time bounds.}

\begin{table}[!t]
\centering
\scriptsize
\caption{{NMPC solver timing for frozen and non-frozen contact-point interpolation
formulations. Timing values are reported in milliseconds for the MATLAB/CasADi/IPOPT
implementation on the Intel Core i7 laptop.}}
\label{tab:table_timing_frozen_vs_nonfrozen}
\resizebox{\textwidth}{!}{%
\begin{tabular}{llcccc}
\hline
Scenario & Method & Mean [ms] & Median [ms] & P95 [ms] & Max [ms] \\
\hline
1 & Frozen    & 325.8 & 298.4 & 434.0  & 2566 \\
1 & NonFrozen & 381.6 & 363.3 & 462.4  & 3745 \\
2 & Frozen    & 357.0 & 302.2 & 411.7  & 2523 \\
2 & NonFrozen & 416.7 & 335.8 & 521.7  & 3730 \\
3 & Frozen    & 340.8 & 269.0 & 576.2  & 2586 \\
3 & NonFrozen & 448.2 & 391.2 & 628.6  & 4020 \\
4 & Frozen    & 408.3 & 301.6 & 1114.0 & 2543 \\
4 & NonFrozen & 463.8 & 393.1 & 619.3  & 3837 \\
5 & Frozen    & 379.0 & 274.5 & 1055.0 & 2997 \\
5 & NonFrozen & 443.6 & 381.0 & 574.1  & 3795 \\
6 & Frozen    & 462.3 & 344.4 & 1226.0 & 3336 \\
6 & NonFrozen & 471.4 & 431.8 & 609.4  & 3820 \\
\hline
\end{tabular}%
}
\end{table}

The timing results in Table~\ref{tab:table_timing_frozen_vs_nonfrozen} show that the frozen formulation reduces the computational burden relative to the non-frozen formulation, with lower mean and median solve times in all reported scenarios and lower 95th-percentile times in several scenarios. Combined with the comparable closed-loop behavior reported in Table~\ref{tab:scenario_metrics}, these results indicate that freezing the spatial contact operators preserves the main control performance {while improving computational tractability and reducing the cost of repeated horizon-wise contact-operator updates.}

{The maximum solve times, however, exceed the sampling time \(T_s=20\) ms in all
tested cases. {These timings characterize the MATLAB/CasADi/IPOPT implementation on the
reported Intel Core i7 laptop as a numerical prototyping benchmark, rather than a hard
real-time embedded implementation.} Achieving strict hard real-time execution would require
a dedicated real-time NMPC implementation, for example through compiled code generation,
real-time iteration SQP, tailored sparse linear algebra, shorter prediction horizons,
smaller reduced-order models, parallelization, or faster embedded/real-time optimization
hardware. Real-time NMPC frameworks such as acados provide code-generation and
real-time-iteration capabilities for fast or embedded optimal-control
applications and represent promising implementation routes for future deployment
\cite{Verschueren2022}.}

\section{Conclusion}
\label{sec:conclusion}

This paper presented a unified modeling-and-control framework for an electric unicycle traversing a
tensioned flexible cable under moving-contact constraints.
A nonlinear finite-element model of the cable, retaining both vertical and lateral transverse dynamics,
was coupled to the unicycle rigid-body roll and pitch dynamics through position-dependent
contact kinematics.
To enable control-oriented prediction, the cable dynamics were reduced using modal truncation while
preserving the coupled nonlinear structure and configuration-dependent inertia of the full system.

Building on this reduced-order nonlinear model, a full nonlinear receding-horizon NMPC problem was
formulated and solved online using CasADi/IPOPT with explicit enforcement of actuator bounds, balance
limits, and cable-span constraints.
To address the dominant computational burden introduced by moving-contact interpolation, a frozen spatial contact-point interpolation strategy was introduced {to reduce the computational burden of nonlinear NMPC for the coupled unicycle--cable system.}

Closed-loop numerical studies demonstrated that the proposed NMPC scheme achieves accurate traversal
along the cable while maintaining attitude stability and respecting actuator limits.
Under plant-only disturbances, constraint tightening, and combined disturbance channels, the controller
maintained closed-loop feasibility in the tested scenarios and attenuated transient deviations,
supporting robustness of the receding-horizon implementation.
{In addition, the frozen-interpolation formulation reduced the NMPC runtimes relative to the non-frozen formulation, indicating improved computational tractability for finite-element-based moving-contact systems.
The present MATLAB/CasADi/IPOPT implementation should be viewed as a numerical prototyping benchmark, while strict hard real-time execution requires a dedicated real-time NMPC implementation.}

Future work will consider experimental validation, extension to more detailed contact models
(e.g., compliance and slip), inclusion of parameter uncertainty
(e.g., tension and damping variations) within robust/stochastic NMPC formulations,
and systematic real-time benchmarking on embedded hardware. Extensions to larger modal
bases and longer spans will also be investigated to characterize performance scaling with
structural discretization and model order. {In addition, future theoretical work will address
formal recursive-feasibility and closed-loop stability guarantees through the construction of
terminal invariant sets and locally admissible terminal feedback laws for the nonlinear
reduced-order unicycle--cable dynamics.}

\appendix
\section{Detailed nonlinear model expressions}
\label{app:detailed_model}

{This appendix collects the expanded algebraic expressions used to construct the
compact nonlinear model in Section~\ref{sec:nonlinear_statespace_ocp}. The main
text retains the Lagrange formulation, generalized-force mapping, compact
state-space representation, and modal-reduction procedure, whereas the
coordinate-wise equations and the detailed matrix/vector entries are reported
here for completeness and reproducibility.}
\subsection{Expanded coordinate-wise equations}
\label{app:expanded_coordinate_equations}

{The following equations correspond to the cable and unicycle coordinate equations
introduced in Section~\ref{sec:model_cable}. They are obtained by applying
Lagrange's equations to the generalized coordinates
\(\begin{bmatrix}\bm q & s & \phi & \theta & \gamma\end{bmatrix}^{\mathsf T}\).}

\paragraph{Cable coordinates.}
\begin{align}
&\Big[\bm M_c+m_u\bm N_v^{\mathsf T}\bm N_v+m_u\bm N_w^{\mathsf T}\bm N_w\Big]\ddot{\bm q}
+\Big[m_u\bm N_v^{\mathsf T}\bm B_v\bm q+m_u\bm N_w^{\mathsf T}\bm B_w\bm q\Big]\ddot{s}
\nonumber\\
&-\Big[\left(m_ur\sin\phi+m_{ab}h\sin\phi\cos\theta\right)\bm N_v^{\mathsf T}
-\left(m_ur\cos\phi+m_{ab}h\cos\phi\cos\theta\right)\bm N_w^{\mathsf T}\Big]\ddot{\phi}
\nonumber\\
&-\Big[\left(m_{ab}h\cos\phi\sin\theta\right)\bm N_v^{\mathsf T}
+\left(m_{ab}h\sin\phi\sin\theta\right)\bm N_w^{\mathsf T}\Big]\ddot{\theta}
+\bm C_c\dot{\bm q}
\nonumber\\
&-\Big[\left(m_ur\cos\phi+m_{ab}h\cos\phi\cos\theta\right)\bm N_v^{\mathsf T}
+\left(m_ur\sin\phi+m_{ab}h\sin\phi\cos\theta\right)\bm N_w^{\mathsf T}\Big]\dot{\phi}^{2}
\nonumber\\
&-\Big[\left(m_{ab}h\cos\phi\cos\theta\right)\bm N_v^{\mathsf T}
+\left(m_{ab}h\sin\phi\cos\theta\right)\bm N_w^{\mathsf T}\Big]\dot{\theta}^{2}
+\Big[2m_u\bm N_v^{\mathsf T}\bm B_v\dot{\bm q}
+2m_u\bm N_w^{\mathsf T}\bm B_w\dot{\bm q}\Big]\dot{s}
\nonumber\\
&+\Big[\left(2m_{ab}h\sin\phi\sin\theta\right)\bm N_v^{\mathsf T}
-\left(2m_{ab}h\cos\phi\sin\theta\right)\bm N_w^{\mathsf T}\Big]\dot{\phi}\dot{\theta}
+\bm K_c\bm q+m_ug\bm N_v^{\mathsf T}
=\bm Q_q .
\label{eq:cable_eom_full}
\end{align}

\paragraph{Axial coordinate.}
\begin{align}
&\Big[m_u\bm B_v\bm q\,\bm N_v+m_u\bm B_w\bm q\,\bm N_w\Big]\ddot{\bm q}
+\Big[m_u+m_u(\bm B_v\bm q)^2+m_u(\bm B_w\bm q)^2+I_{wz}/r^2\Big]\ddot{s}
\nonumber\\
&-\Big[\left(m_ur\sin\phi+m_{ab}h\sin\phi\cos\theta\right)\bm B_v\bm q
-\left(m_ur\cos\phi+m_{ab}h\cos\phi\cos\theta\right)\bm B_w\bm q\Big]\ddot{\phi}
\nonumber\\
&-\Big[m_{ab}h\cos\theta
+\left(m_{ab}h\cos\phi\sin\theta\right)\bm B_v\bm q
+\left(m_{ab}h\sin\phi\sin\theta\right)\bm B_w\bm q\Big]\ddot{\theta}
\nonumber\\
&-\Big[\left(m_ur\cos\phi+m_{ab}h\cos\phi\cos\theta\right)\bm B_v\bm q
+\left(m_ur\sin\phi+m_{ab}h\sin\phi\cos\theta\right)\bm B_w\bm q\Big]\dot{\phi}^{2}
\nonumber\\
&+\Big[m_{ab}h\sin\theta
-\left(m_{ab}h\cos\phi\cos\theta\right)\bm B_v\bm q
-\left(m_{ab}h\sin\phi\cos\theta\right)\bm B_w\bm q\Big]\dot{\theta}^{2}
\nonumber\\
&+\Big[2m_u(\bm B_v\bm q)\bm B_v\dot{\bm q}
+2m_u(\bm B_w\bm q)\bm B_w\dot{\bm q}\Big]\dot{s}
\nonumber\\
&+\Big[\left(2m_{ab}h\sin\phi\sin\theta\right)\bm B_v\bm q
-\left(2m_{ab}h\cos\phi\sin\theta\right)\bm B_w\bm q\Big]\dot{\phi}\dot{\theta}
+m_ug\bm B_v\bm q
=Q_s .
\label{eq:s_eom_full}
\end{align}

\paragraph{Roll coordinate.}
\begin{align}
&-\Big[\left(m_ur\sin\phi+m_{ab}h\sin\phi\cos\theta\right)\bm N_v
-\left(m_ur\cos\phi+m_{ab}h\cos\phi\cos\theta\right)\bm N_w\Big]\ddot{\bm q}
\nonumber\\
&-\Big[\left(m_ur\sin\phi+m_{ab}h\sin\phi\cos\theta\right)\bm B_v\bm q
-\left(m_ur\cos\phi+m_{ab}h\cos\phi\cos\theta\right)\bm B_w\bm q\Big]\ddot{s}
\nonumber\\
&+\Big[m_ur^2+I_{wx}+I_{ax}
+\left(I_{bx}+m_{ab}h^2\right)\cos^2\theta
+I_{by}\sin^2\theta
+2m_{ab}hr\cos\theta\Big]\ddot{\phi}
+I_{ax}\ddot{\gamma}
\nonumber\\
&-\Big[2\left(m_ur\sin\phi+m_{ab}h\sin\phi\cos\theta\right)\bm B_v\dot{\bm q}
-2\left(m_ur\cos\phi+m_{ab}h\cos\phi\cos\theta\right)\bm B_w\dot{\bm q}\Big]\dot{s}
\nonumber\\
&-\Big[2m_{ab}hr\sin\theta
+2\left(m_{ab}h^2+I_{bx}-I_{by}\right)\sin\theta\cos\theta\Big]\dot{\phi}\dot{\theta}
\nonumber\\
&-m_ugr\sin\phi-m_{ab}gh\sin\phi\cos\theta
=Q_{\phi}.
\label{eq:phi_eom_full}
\end{align}

\paragraph{Pitch coordinate.}
\begin{align}
&-\Big[\left(m_{ab}h\cos\phi\sin\theta\right)\bm N_v
+\left(m_{ab}h\sin\phi\sin\theta\right)\bm N_w\Big]\ddot{\bm q}
\nonumber\\
&-\Big[m_{ab}h\cos\theta
+\left(m_{ab}h\cos\phi\sin\theta\right)\bm B_v\bm q
+\left(m_{ab}h\sin\phi\sin\theta\right)\bm B_w\bm q\Big]\ddot{s}
\nonumber\\
&+\Big[m_{ab}h^2+I_{bz}+I_{az}\Big]\ddot{\theta}
\nonumber\\
&-\Big[2\left(m_{ab}h\cos\phi\sin\theta\right)\bm B_v\dot{\bm q}
+2\left(m_{ab}h\sin\phi\sin\theta\right)\bm B_w\dot{\bm q}\Big]\dot{s}
\nonumber\\
&+\Big[m_{ab}hr\sin\theta
+\left(m_{ab}h^2+I_{bx}-I_{by}\right)\sin\theta\cos\theta\Big]\dot{\phi}^{2}
-m_{ab}gh\cos\phi\sin\theta
=Q_{\theta}.
\label{eq:theta_eom_full}
\end{align}

\subsection{Mass-matrix and drift-vector entries}
\label{app:mass_drift_entries}

{The block-partitioned mass matrix in Eq.~\eqref{eq:M_block} has the following
nonzero entries.}

\begin{align}
\bm M_{qq}
&=\bm M_c
+m_u\,\bm N_v^{\mathsf T}\bm N_v
+m_u\,\bm N_w^{\mathsf T}\bm N_w,\label{eq:Mqq}\\
\bm M_{qs}
&=m_u\,\bm N_v^{\mathsf T}\nu_v
+m_u\,\bm N_w^{\mathsf T}\nu_w,\qquad
\bm M_{sq}=\bm M_{qs}^{\mathsf T},\label{eq:Mqs}\\
\bm M_{q\phi}
&=
-\Big[
\big(m_u r\sin\phi+m_{ab}h\sin\phi\cos\theta\big)\bm N_v^{\mathsf T}
-\big(m_u r\cos\phi+m_{ab}h\cos\phi\cos\theta\big)\bm N_w^{\mathsf T}
\Big],\qquad
\bm M_{\phi q}=\bm M_{q\phi}^{\mathsf T},\label{eq:Mqphi}\\
\bm M_{q\theta}
&=
-\Big[
\big(m_{ab}h\cos\phi\sin\theta\big)\bm N_v^{\mathsf T}
+\big(m_{ab}h\sin\phi\sin\theta\big)\bm N_w^{\mathsf T}
\Big],\qquad
\bm M_{\theta q}=\bm M_{q\theta}^{\mathsf T},\label{eq:Mqtheta}\\
M_{ss}
&=m_u+m_u\nu_v^2+m_u\nu_w^2+\frac{I_{wz}}{r^2},\label{eq:Mss}\\
M_{s\phi}
&=
-\Big[
\big(m_u r\sin\phi+m_{ab}h\sin\phi\cos\theta\big)\nu_v
-\big(m_u r\cos\phi+m_{ab}h\cos\phi\cos\theta\big)\nu_w
\Big],\qquad
M_{\phi s}=M_{s\phi},\label{eq:Msphi}\\
M_{s\theta}
&=
-\Big[
m_{ab}h\cos\theta
+\big(m_{ab}h\cos\phi\sin\theta\big)\nu_v
+\big(m_{ab}h\sin\phi\sin\theta\big)\nu_w
\Big],\qquad
M_{\theta s}=M_{s\theta},\label{eq:Mstheta}\\
M_{\phi\phi}
&=
m_u r^2+I_{wx}+I_{ax}
+\big(I_{bx}+m_{ab}h^2\big)\cos^2\theta
+I_{by}\sin^2\theta
+2m_{ab}hr\cos\theta,\label{eq:Mphiphi}\\
M_{\phi\gamma}
&=I_{ax},\qquad
M_{\gamma\phi}=I_{ax},\label{eq:Mphigamma}\\
M_{\theta\theta}
&=m_{ab}h^2+I_{bz}+I_{az},\label{eq:Mthetatheta}\\
M_{\gamma\gamma}
&=I_{ax}.\label{eq:Mgammagamma}
\end{align}
All unspecified blocks are identically zero.

{The nonlinear drift vector in Eq.~\eqref{eq:d_partition} is composed of the
following coordinate-wise components.}

\paragraph{Cable coordinates.}
\begin{equation}
{
\begin{aligned}
\bm d_q
={}&\bm C_c\dot{\bm q}
-\Big[
\big(m_u r\cos\phi+m_{ab}h\cos\phi\cos\theta\big)\bm N_v^{\mathsf T}
+\big(m_u r\sin\phi+m_{ab}h\sin\phi\cos\theta\big)\bm N_w^{\mathsf T}
\Big]\dot\phi^{2}\\
&-\Big[
\big(m_{ab}h\cos\phi\cos\theta\big)\bm N_v^{\mathsf T}
+\big(m_{ab}h\sin\phi\cos\theta\big)\bm N_w^{\mathsf T}
\Big]\dot\theta^{2}\\
&+\Big[2m_u\bm N_v^{\mathsf T}\bm B_v\dot{\bm q}
+2m_u\bm N_w^{\mathsf T}\bm B_w\dot{\bm q}\Big]\dot s\\
&+\Big[
\big(2m_{ab}h\sin\phi\sin\theta\big)\bm N_v^{\mathsf T}
-\big(2m_{ab}h\cos\phi\sin\theta\big)\bm N_w^{\mathsf T}
\Big]\dot\phi\dot\theta\\
&+\bm K_c\bm q+m_u g\,\bm N_v^{\mathsf T}.
\end{aligned}}
\label{eq:dq}
\end{equation}

\paragraph{Axial coordinate.}
\begin{equation}
{
\begin{aligned}
h_s
={}&
-\Big[
\big(m_u r\cos\phi+m_{ab}h\cos\phi\cos\theta\big)\nu_v
+\big(m_u r\sin\phi+m_{ab}h\sin\phi\cos\theta\big)\nu_w
\Big]\dot\phi^{2}\\
&+\Big[
m_{ab}h\sin\theta
-\big(m_{ab}h\cos\phi\cos\theta\big)\nu_v
-\big(m_{ab}h\sin\phi\cos\theta\big)\nu_w
\Big]\dot\theta^{2}\\
&+\Big[2m_u\nu_v\,\dot\nu_v+2m_u\nu_w\,\dot\nu_w\Big]\dot s\\
&+\Big[
\big(2m_{ab}h\sin\phi\sin\theta\big)\nu_v
-\big(2m_{ab}h\cos\phi\sin\theta\big)\nu_w
\Big]\dot\phi\dot\theta\\
&+m_u g\,\nu_v .
\end{aligned}}
\label{eq:hs}
\end{equation}

\paragraph{Roll coordinate.}
\begin{equation}
{
\begin{aligned}
h_{\phi}
={}&
-\Big[2\big(m_u r\sin\phi+m_{ab}h\sin\phi\cos\theta\big)\dot\nu_v
-2\big(m_u r\cos\phi+m_{ab}h\cos\phi\cos\theta\big)\dot\nu_w\Big]\dot s\\
&-\Big[2m_{ab}hr\sin\theta
+2\big(m_{ab}h^2+I_{bx}-I_{by}\big)\sin\theta\cos\theta\Big]\dot\phi\dot\theta\\
&-m_u g r\sin\phi-m_{ab}g h\sin\phi\cos\theta .
\end{aligned}}
\label{eq:hphi}
\end{equation}

\paragraph{Pitch coordinate.}
\begin{equation}
{
\begin{aligned}
h_{\theta}
={}&
-\Big[2\big(m_{ab}h\cos\phi\sin\theta\big)\dot\nu_v
+2\big(m_{ab}h\sin\phi\sin\theta\big)\dot\nu_w\Big]\dot s\\
&+\Big[m_{ab}hr\sin\theta
+\big(m_{ab}h^2+I_{bx}-I_{by}\big)\sin\theta\cos\theta\Big]\dot\phi^{2}
-m_{ab}g h\cos\phi\sin\theta .
\end{aligned}}
\label{eq:htheta}
\end{equation}

\paragraph{Counter-arm coordinate.}
\begin{equation}
h_{\gamma}=0.
\label{eq:hgamma}
\end{equation}
\section*{Declarations}

\textbf{Funding:} No funding was received for this work.

\textbf{Conflict of interest:} The authors declare no conflicts of interest.

\textbf{Ethics approval:} Not applicable.

\textbf{Consent to participate:} Not applicable.

\textbf{Consent to publish:} Not applicable.

\textbf{Data availability:} No datasets were generated or analyzed during this study.

\textbf{Code availability:} Available upon reasonable request.

\textbf{Author contributions:} All authors contributed to the work.

\textbf{Generative AI use in manuscript preparation:} During the preparation of this work, the authors used ChatGPT to improve the language and readability. The authors reviewed and edited the output as needed and take full responsibility for the content of the published article.

\bibliographystyle{elsarticle-num}
\bibliography{EUonCable}

\end{document}